\documentclass[useAMS,usenatbib,usegraphicx]{mn2e}
\usepackage{epsfig}
\usepackage{amsmath} 
\usepackage{rotating}           
\usepackage{color}     
\usepackage{graphicx}
\usepackage{times}
\usepackage{upgreek} 
\usepackage{multicol}

\def\lapp{\ifmmode\stackrel{<}{_{\sim}}\else$\stackrel{<}{_{\sim}}$\fi}
\def\gapp{\ifmmode\stackrel{>}{_{\sim}}\else$\stackrel{>}{_{\sim}}$\fi}

\title[Redshifts of radio sources in the Milliquas]{Redshifts of radio sources in the Million Quasars Catalogue from machine learning}

\author[S. J. Curran, J. P. Moss \& Y. C. Perrott]{S. J. Curran\thanks{Stephen.Curran@vuw.ac.nz}, J. P. Moss and Y. C. Perrott\\
School of Chemical and Physical Sciences, Victoria University of Wellington, PO Box 600, Wellington 6140, New Zealand}

\begin{document}

 \date{Accepted ---. Received ---; in original form ---}

\pagerange{\pageref{firstpage}--\pageref{lastpage}} \pubyear{2022}

\maketitle
\label{firstpage}
\begin{abstract}
  With the aim of using machine learning techniques to obtain photometric redshifts based upon a source's radio spectrum
  alone, we have extracted the radio sources from the {\em Million Quasars Catalogue}. Of these, 44\,119 have a
  spectroscopic redshift, required for model validation, and for which photometry could be obtained.  Using the radio
  spectral properties as features, we fail to find a model which can reliably predict the redshifts, although there is
  the suggestion that the models improve with the size of the training sample.  Using the near-infrared--optical--ultraviolet bands  magnitudes, we obtain reliable predictions based on the 12\,503 radio sources which
  have all of the required photometry. From the 80:20 training--validation split, this gives only 2501 validation
  sources, although training the sample upon our previous SDSS model gives comparable results for all 12\,503
  sources. This makes us confident that SkyMapper, which will survey southern sky in the $u, v, g, r, i, z$ bands, can
  be used to predict the redshifts of radio sources detected with the {\em Square Kilometre Array}. By using machine
  learning to impute the magnitudes missing from much of the sample, we can predict the redshifts for 32\,698 sources,
  an increase from 28\% to 74\% of the sample, at the cost of increasing the outlier fraction by a factor of 1.4. While
  the ``optical'' band data prove successful, at this stage we cannot rule out the
  possibility of a radio photometric redshift, given sufficient data which may be necessary to overcome the relatively
  featureless radio spectra. 
\end{abstract}  
\begin{keywords}
{techniques: photometric  -- methods: statistical --  galaxies: active --  galaxies: photometry -- infrared: galaxies -- ultraviolet: galaxies}
\end{keywords}

\section{Introduction} 
\label{intro}

There is currently much interest in using machine learning methods to determine the redshifts of distant sources from their
photometry. Since forthcoming continuum surveys on the next generation of telescopes are expected yield vast number of
new detections, obtaining spectroscopic redshifts for each of these will be too observationally expensive.
Of particular interest (to us), is the redshifts of the new sources detected in the radio band with the {\em
  Square Kilometre Array} (SKA), where even just the {\em Australian Square Kilometre Array Pathfinder} is expected to 
yield 70 million radio detections (\citealt{nha+11}). Being able to determine the distance to these will greatly
increase the scientific value of the surveys. Furthermore, photometric redshifts can be obtained for objects significantly 
fainter than possible with optical spectroscopy, thus not being as susceptible
to bias against the more dust obscured objects, most likely to host cold,
star-forming gas, detected through the absorption of the 21-cm transition of hydrogen and the rotational transitions of molecules
\citep{cwc+11,cw12,cur21b}. 

Finding a photometric redshift based upon the radio spectrum alone has so far proven elusive \citep{maj15,nsl+19},
most likely due to the relatively featureless spectra, whereas machine learning techniques
using the near-infrared--optical--ultra-violet photometry have proven successful \citep{bmh+12,bcd+13,cma+21,nbp+21}. 
These are usually trained and validated upon the same sample, typically  from the {\em Sloan Digital Sky Survey} (SDSS),
although \citet{cur20,cmp21} have demonstrated that an SDSS trained model can reliably predict the redshifts of quasars
from external catalogues of radio-selected sources. 

\citet{tdss20} have used the radio fluxes at 151, 178 and 2700~MHz, combined with imaging of the radio lobes, to predict the redshifts
of nine sources, although it is not clear how this will scale to large samples where the emission may be spatially unresolved.
Ideally, we wish to be able to predict the redshifts from the radio spectral energy distribution (SED) alone,
 although as discussed by \citet{cm19},
there is a dearth of large radio catalogues for which we also have accurate spectroscopic redshifts \citep{msc+15}, required to enable 
validation of the predictions.  

In this paper, we therefore extract the radio sources in the 1\,573\,822 strong  {\em Million Quasars Catalogue}  
(Milliquas, \citealt{fle21}) and explore the potential of this sample in the prediction of radio photometric redshifts.
We also test the accuracy of a model constructed from the near-infrared--optical--ultra-violet  of the Milliquas in
predicting redshifts for radio-selected sources.

\section{The data} 

\subsection{The sample}

The Milliquas comprises 1\,573\,822 quasi-stellar objects (QSOs)\footnote{As usual, we shall refer to the optically
selected objects as QSOs and the radio selected as quasars.} compiled from the literature,
including SDSS QSOs up to  Data Release 16  (\citealt{fle15, fle21}). Of these, there are 136\,076
with a radio association, of which  44\,495 have a spectroscopic redshift.

\subsection{Photometry extraction}
\label{pe}

Similarly to what we did in \citet{cmp21}, each of the final 44\,495 sources was
matched to a source in the {\em NASA/IPAC Extragalactic  Database} (NED). 
Matches were found for 44\,119, for which we scraped all of the photometry, including the radio-band data. 
Since the  $W2, W1, z,i,r,g,u,NUV,FUV$ photometry have a proven track record in the use of 
machine learning to predict redshifts  (Sect.~\ref{intro}),
we used the NED names to query  the {\em Wide-Field Infrared Survey Explorer} (WISE), the
{\em Two Micron All Sky Survey} (2MASS, \citealt{scs+06}) databases  for the near-infrared (NIR)  
and the GALEX database \citep{bst17} for ultra-violet (UV) photometry.

\subsection{Photometry fitting}

\subsubsection{Radio fitting and features}
\label{rfit}

The combination of NIR, optical and UV colours have been very successful when used as features for machine learning
models and artificial neural networks, with the raw magnitudes appearing to perform as well as the colour indices
\citep{cmp21,cur22}.  We therefore use the analogous radio features here, specifically the flux density at various
frequencies. As per the NIR--optical--UV fitting (see Sect.~\ref{opt_fit}), we select several commonly observed bands
and average the fluxes within each of these for each source, specifically at 70~MHz (over 70--80~MHz), 150~MHz
(150--160~MHz), 400~MHz (365--408~MHz), 700~MHz (635--750~MHz), 1~GHz (960--1100~MHz), 1.4~GHz (1.3--1.5~GHz), 2.7~GHz
(2.6--2.7~GHz), 5~GHz (4.7--5.0~GHz), 8.7~GHz (7.9--8.9~GHz), 15~GHz (14--16~GHz) and 20~GHz (20--24~GHz).

\begin{figure}
\centering \includegraphics[angle=-90,scale=0.52]{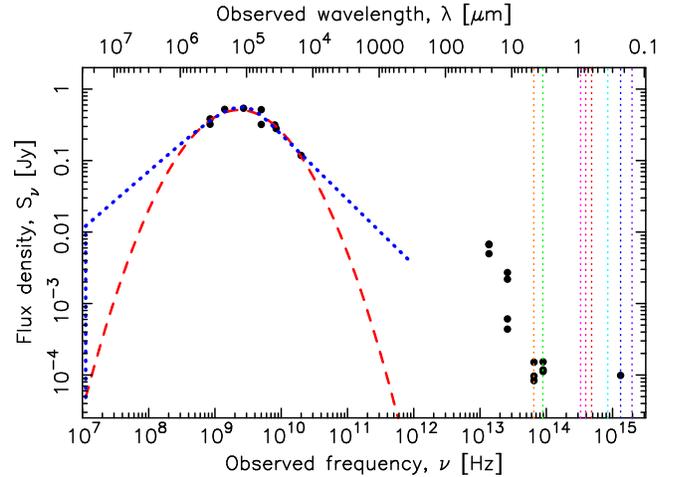}
\caption{Example of the radio-band fitting to the photometry. The red broken curve shows
the second order polynomial fitting, giving $\nu_{\text{TO}}$, the blue dotted curve the \citeauthor{ssd+98} fit, giving
$\nu_{\text{peak}}$, $\alpha_{\text{thin}}$ (to the left of $\nu_{\text{peak}}$) and 
$\alpha_{\text{thick}}$ (to the right of $\nu_{\text{peak}}$). The dotted vertical lines show the location of the 
$W2, W1, z,i,r,g,u,NUV,FUV$ bands.}
\label{SED}
\end{figure}

In order to test for a turnover in the SED,  we also fit a second order polynomial to the radio photometry (see \citealt{cwsb12}).
This was used to determine the spectral index at 1.4~GHz, $\alpha_{1.4~\text{GHz}}$, as well as providing the initial parameters
to fit a function of the form \citep{ssd+98}
\[
S = \frac{S_{\text{peak}}}{1 - 1/e}\left(\frac{\nu}{\nu_{\text{peak}}}\right)^{\alpha_{\text{thick}}}\left(1-\exp\left\{-\left(\frac{\nu}{\nu_{\text{peak}}}\right)^{\alpha_{\text{thin}}- \alpha_{\text{thick}}}\right\}\right),
\]
where $S_{\text{peak}}$ is the flux density at the turnover frequency, $\nu_{\text{peak}}$,  $\alpha_{\text{thick}}$ the spectral index of optically thick part of spectrum and $\alpha_{\text{thin}}$ the spectral index of optically thin part of spectrum (Fig.~\ref{SED}). 

\begin{figure*}
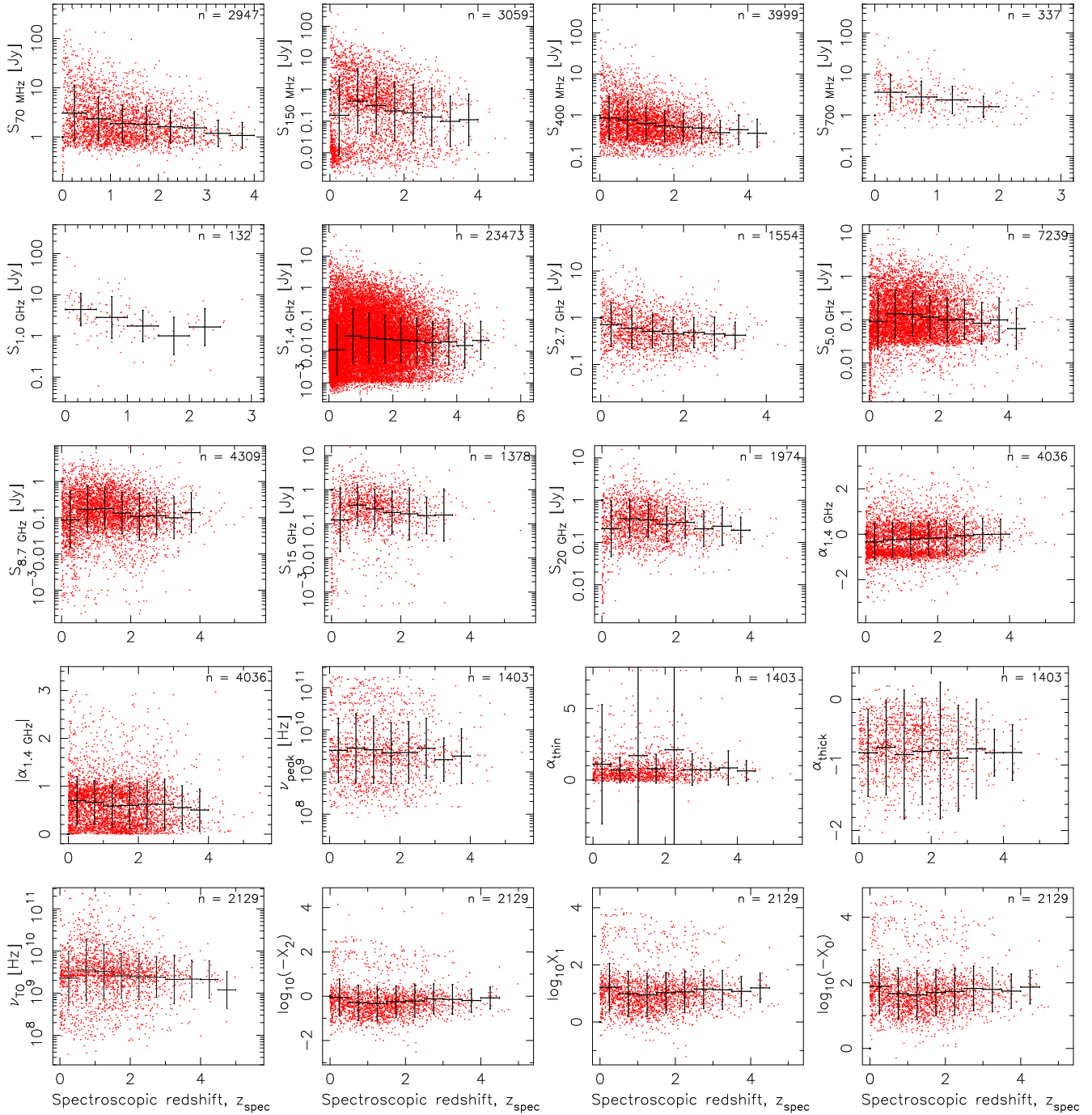

\centering \includegraphics[angle=-90,scale=0.26]{S_70-z_294.eps}
\centering \includegraphics[angle=-90,scale=0.26]{S_150-z_278.eps}
\centering \includegraphics[angle=-90,scale=0.26]{S_400-z_399.eps}
\centering \includegraphics[angle=-90,scale=0.26]{S_700-z_84.eps}
\centering \includegraphics[angle=-90,scale=0.26]{S_1-z_26.eps}
\centering \includegraphics[angle=-90,scale=0.26]{S_1p4-z_2347.eps}
\centering \includegraphics[angle=-90,scale=0.26]{S_2p7-z_155.eps}
\centering \includegraphics[angle=-90,scale=0.26]{S_5-z_723.eps}
\centering \includegraphics[angle=-90,scale=0.26]{S_8p7-z_430.eps}
\centering \includegraphics[angle=-90,scale=0.26]{S_15-z_137.eps}
\centering \includegraphics[angle=-90,scale=0.26]{S_20-z_197.eps}
\centering \includegraphics[angle=-90,scale=0.26]{SI_z_403.eps}
\centering \includegraphics[angle=-90,scale=0.26]{SI-abs_z_403.eps}
\centering \includegraphics[angle=-90,scale=0.26]{TO_ssd+98-z_140.eps}
\centering \includegraphics[angle=-90,scale=0.26]{thin-z_140.eps}
\centering \includegraphics[angle=-90,scale=0.26]{thick_z_140.eps}
\centering \includegraphics[angle=-90,scale=0.26]{TO_poly-z_212.eps}
\centering \includegraphics[angle=-90,scale=0.26]{X2_z_212.eps}
\centering \includegraphics[angle=-90,scale=0.26]{X1_z_212.eps}
\centering \includegraphics[angle=-90,scale=0.26]{X0_z_212.eps}
\caption{Distributions of potential radio features with redshift. The binning is for equal redshift spacing, with the error 
bars on the ordinate showing the mean value $\pm1\sigma$. }
\label{radio_plots}
\end{figure*}

We summarise these potential features in Table~\ref{radio_feat},
\begin{table*}  
\begin{minipage}{135mm}
\centering
 \caption{Potential features in the classification of  the radio spectra (see Fig.~\ref{radio_plots}).}
\begin{tabular}{@{}l  l @{}}
\hline
\smallskip
Feature(s) & Description \\
\hline
$\log_{10} S_{\text{70 MHz}}$ to $\log_{10} S_{\text{20 GHz}} $  & Log flux densities at 70, 150, 400 \& 700 ~MHz and 1.0, 1.4, 2.7, 5.0. 8.7, 15 \&  20~GHz\\
\hline
\multicolumn{2}{c}{Polynomial fit}\\
\hline
$\alpha_{1.4~\text{GHz}}$ & Spectral index at 1.4 GHz\\
$X_2, X_1, X_0$ & Coefficients of the polynomial fit,  $\log_{10} S = X_2\log_{10}\nu^2 + X_1\log_{10} \nu + \log_{10}X_0$ \\ 
$\log_{10}\nu_{\text{TO}}$  & Log turnover frequency of the fit, i.e.  where  $dS/d\nu = -X_1/2X_2= 0$\\ 
$\log_{10} S_{\text{TO}}$  & Log flux density at turnover frequency\\
\hline
\multicolumn{2}{c}{Additional parameters from \citeauthor{ssd+98} fit}\\
\hline
$\log_{10} \nu_{\text{peak}}$ & Log turnover frequency of the fit\\
$\log_{10} S_{\text{peak}}$ & Log flux density at turnover frequency\\
$\alpha_{\text{thin}}$ & Spectral index of optically thin part of spectrum\\
$\alpha_{\text{thick}}$ & Spectral index of optically thick part of spectrum\\
\hline
\end{tabular}
\label{radio_feat}  
\end{minipage}
\end{table*} 
and show their redshift distributions in Fig.~\ref{radio_plots}.
From these (Table~\ref{z_stats}), we 
\begin{table}  
\begin{minipage}{65mm}
\centering
 \caption{The feature redshift correlations (Fig.~\ref{radio_plots}). The number of sources is followed by 
the Kendall-tau test probability of the observed correlation arising by chance and the significance of the
correlation ($Z$-value), assuming Gaussian statistics.}
\begin{tabular}{@{}l  r c r @{}}
\hline
\smallskip
Feature &  $n$ & $P(\tau)$  &  $Z(\tau)$\\ 
\hline
$\log_{10} S_{\text{70 MHz}}$  & 2947 & $1.15\times10^{-26} $& $10.60\sigma$ \\ 
$\log_{10} S_{\text{150 MHz}}$  & 3059 & 0.319 & $0.97\sigma$ \\ 
$\log_{10} S_{\text{400 MHz}}$  & 3999 & $2.27\times10^{-31} $& $11.65\sigma$ \\ 
$\log_{10} S_{\text{700 MHz}}$  & 337 & $1.51\times10^{-8} $& $5.66\sigma$ \\ 
$\log_{10} S_{\text{1.0 GHz}}$  & 132 & $6.25\times10^{-6} $& $4.25\sigma$ \\ 
$\log_{10} S_{\text{1.4 GHz}}$  & 23\,473& $7.16\times10^{-141}$&  $19.70\sigma$ \\ 
$\log_{10} S_{\text{2.7 GHz}}$  & 1554 & $1.29\times10^{-11} $& $6.77\sigma$ \\ 
$\log_{10} S_{\text{5.0 GHz}}$  & 7239 & 0.379 & $0.88\sigma$ \\ 
$\log_{10} S_{\text{8.7 GHz}}$  & 4309 & 0.963 &  $0.05\sigma$ \\ 
$\log_{10} S_{\text{15 GHz}}$ & 1378 & 0.721 & $0.36\sigma$ \\ 
$\log_{10} S_{\text{20 GHz}}$ & 1974& 0.269 & $1.11\sigma$ \\ 
$\alpha_{1.4~\text{GHz}}$ & 4036 & $4.45\times10^{-15} $& $7.84\sigma$ \\ 
$|\alpha_{1.4~\text{GHz}}|$ & 4036 & $6.36\times10^{-9} $& $5.81\sigma$ \\ 
$\log_{10} \nu_{\text{peak}}$ & 1403 & 0.342 & $0.95\sigma$ \\ 
$\log_{10} S_{\text{peak}}$ & 1403 & 0.017& $2.40\sigma$ \\ 
$\alpha_{\text{thin}}$ & 1403 & 0.190 & $1.31\sigma$ \\ 
$\alpha_{\text{thick}}$ & 1403 & 0.960 & $0.05\sigma$ \\ 
$\log_{10}\nu_{\text{TO}}$     & 2129 & 0.125 & $1.59\sigma$ \\
$\log_{10}S_{\text{TO}}$     & 2129 & 0.230 & $1.20\sigma$ \\
$\log_{10} (-X_2)$  & 2129&  0.754& $0.31\sigma$ \\
$\log_{10} X_1$  &2129&  0.772& $0.29\sigma$ \\
$\log_{10} (-X_0)$  & 2129&  0.785& $0.27\sigma$ \\
\hline
\end{tabular}
\label{z_stats}  
\end{minipage}
\end{table} 
see that, as expected, the flux densities are generally anti-correlated with the redshift.\footnote{Those of low significance
appear to be due to a low redshift bump, negating the correlation apparent at $z\gapp1$.} Apart from these,
the only other parameter strongly correlated with the redshift is the spectral index, $\alpha_{1.4~\text{GHz}}$.
A positive correlation between spectral index and radio luminosity (and therefore redshift) has been previously noted 
for 69 radio galaxies and quasars by \citet{bkg72}, with \citet{ak98} finding a positive correlation for 15 $z>2$
radio galaxies. 
Also, like \citet{ob97}, we find no correlation with the turnover frequency (cf. \citealt{men83}). 
Finally, in Table~\ref{intersect_all}, we list the number of  features common to the sources,
\begin{table}  
\begin{minipage}{85mm}
\centering
 \caption{Features common to  the sources in (column) descending order of sample size (Table~\ref{z_stats}).
$n_{\text{int}}$ gives the accumulated intersection of features, $\log_{10} S_{\text{1.4 GHz}} \cap \log_{10} S_{\text{5.0 GHz}}$ in the
second row, followed in $\log_{10} S_{\text{1.4 GHz}} \cap \log_{10} S_{\text{5.0 GHz}} \cap \alpha_{1.4~\text{GHz}}$  in the third
row, etc.}
\begin{tabular}{@{}l  r r | l  rr @{}}
\hline
\smallskip
Feature &  $n$ & $n_{\text{int}}$ & Feature &  $n$ & $n_{\text{int}}$ \\ 
\hline
\multicolumn{6}{c}{Whole sample}\\
\hline
$\log_{10} S_{\text{1.4 GHz}}$ & 23\,473&  23\,473&Polynomial fit & 2129 & 76\\               
$\log_{10} S_{\text{5.0 GHz}}$  & 7239 & 6808&     $\log_{10} S_{\text{20 GHz}}$ & 1974& 60\\ 
$\log_{10} S_{\text{8.7 GHz}}$  & 4309 & 3691&     $\log_{10} S_{\text{2.7 GHz}}$ & 1554 &45\\
$\alpha_{1.4~\text{GHz}}$ &  4036 & 2058&          \citeauthor{ssd+98} fit & 1403  & 42 \\    
$\log_{10} S_{\text{400 MHz}}$  &3999 &  1001&     $\log_{10} S_{\text{15 GHz}}$ & 1378 & 39\\
$\log_{10} S_{\text{150 MHz}}$  & 3059 & 351&      $\log_{10} S_{\text{700 MHz}}$  & 337 & 25\\
$\log_{10} S_{\text{70 MHz}}$ & 2947 & 253 &      $\log_{10} S_{\text{1.0 GHz}}$ & 132 & 12\\
\hline
\multicolumn{6}{c}{Most populous with polynomial fit ({\sc poly} sample)}\\
\hline
Polynomial fit & 2129 & 2129 & $\log_{10} S_{\text{8.7 GHz}}$  & 4309 & 1421\\
$\log_{10} S_{\text{1.4 GHz}}$ & 23\,473& 1976 & $\alpha_{1.4~\text{GHz}}$ &  4036 &1416\\
$\log_{10} S_{\text{5.0 GHz}}$ &  7239 & 1744 & $\log_{10} S_{\text{400 MHz}}$ & 3999 & 559\\
\hline
\multicolumn{6}{c}{Most populous with  \citeauthor{ssd+98} fit ({\sc snellen} sample)}\\
\hline
\citeauthor{ssd+98} fit &  1403& 1403 & $\log_{10} S_{\text{8.7 GHz}}$  & 4309 & 991\\
$\log_{10} S_{\text{1.4 GHz}}$ & 23\,473& 1289 & $\alpha_{1.4~\text{GHz}}$ &  4036 &990\\
$\log_{10} S_{\text{5.0 GHz}}$ &  7239 & 1163 & $\log_{10} S_{\text{400 MHz}}$ & 3999 & 478\\
\hline
\multicolumn{6}{c}{Flux densities only ({\sc flux} sample)} \\
\hline
$\log_{10} S_{\text{1.4 GHz}}$ & 23\,473 & 23\,473 & $\log_{10} S_{\text{20 GHz}}$ & 1974& 329\\
 $\log_{10} S_{\text{5.0 GHz}}$ &  7239 & 6808 &     $\log_{10} S_{\text{2.7 GHz}}$ & 1554 &233\\
$\log_{10} S_{\text{8.7 GHz}}$  & 4309 & 3691 &   $\log_{10} S_{\text{15 GHz}}$ & 1378 & 137\\
$\log_{10} S_{\text{400 MHz}}$ & 3999 & 2037 & $\log_{10} S_{\text{700 MHz}}$  & 337 & 83\\
$\log_{10} S_{\text{150 MHz}}$  & 3059 & 716 & $\log_{10} S_{\text{1.0 GHz}}$ & 132 & 48 \\
$\log_{10} S_{\text{70 MHz}}$ & 2947 & 549 & & & \\
\hline
\end{tabular}
\flushleft 
{Notes: {\em Polynomial fit} comprises the features $\log_{10}\nu_{\text{TO}}$ , $\log_{10}S_{\text{TO}}$, $\log_{10} (-X_2)$, $\log_{10} X_1$   and $\log_{10} (-X_0)$. {\em \citeauthor{ssd+98} fit} comprises the features $\log_{10} \nu_{\text{peak}}$, $\log_{10} S_{\text{peak}}$, $\alpha_{\text{thin}}$ and $\alpha_{\text{thick}}$.}
\label{intersect_all}  
\end{minipage}
\end{table} 
from which we see only 12 of the 44\,119 have all  features.

\subsubsection{NIR--optical--UV fitting}
\label{opt_fit}

As before \citep{cmp21}, for the optical-band photometry the PSF flux densities associated with the AB magnitudes of the 
{\em Sloan Digital Sky Survey} (SDSS) were used. We chose these as they are the most extensive available and 
are the standard choice in using machine learning to obtain photometric redshifts
(e.g. \citealt{rws+01,wrs+04,mhp+12,hdzz16}).
In order to be applicable to samples for which the SDSS, WISE and GALEX photometry may not be directly available, including the 
current sample, we collected data which fall within $\Delta\log_{10}\nu = \pm0.05$ 
of the central frequency of each of the bands. Within each band the fluxes were then averaged
before being converted to a magnitude.\footnote{For GALEX this was via 
$M = -2.5(\log_{10}S_{\nu}-3.56)$, where $S_{\nu}$ is the specific flux density in Jy (http://galex.stsci.edu/gr6/).}
Of the 44\,119 radio sources, there were 12\,503 which had all nine of the required magnitudes.

\section{Methodology}

\subsection{Machine Learning}

We tested three machine learning algorithms, in addition to an {\em Artificial Neural Network} (ANN) algorithm.
For the machine learning we used {\em $k$-Nearest Neighbour Regressor} (kNN),  {\em Support Vector Regression} (SVR)  and 
{\em Decision Tree Regression} (DTR), which were initally tested with 
the default hyperparameters (see https://scikit-learn.org/stable/).  After experimenting with these, we settled on:
\begin{itemize}
\item kNN -- uniform weighting and $k \approx10-25$ nearest neighbours, 
\item SVR -- $C\approx50$, $\gamma =0.1$ and $\epsilon=0.1$ (cf. the default parameters $C=10$, $\gamma =$\,{\tt scale} and  $\epsilon=0.1$)\footnote{https://scikit-learn.org/stable/modules/generated/sklearn.svm.SVR.html}, 
\item DTR -- maximum depth of 4 -- 12,
\end{itemize}
depending on the dataset (Sect.~\ref{rpr}). Note that, perhaps due to the small sample sizes, the algorithms were
fairly insensitive to the hyperparameters (especially over many trials, see Table~\ref{ML_TO}).
Also, when close to what appeared to be the optimum model,  tweaking the models could move the 
metrics in opposite
directions, e.g. an increase in the regression coefficient together with an increase in the spread between the
predicted and measured redshifts (see below).

\begin{figure}
\centering\includegraphics[angle=0,scale=0.52]{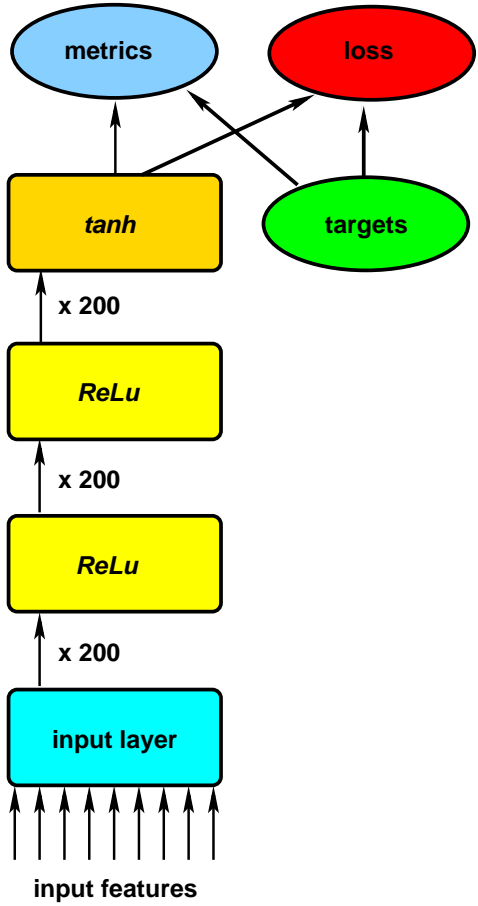} 
\caption{Architecture of the neural network used in the artificial neutral network.} 
\label{fig:nn_architecture}
\end{figure}

Although samples which included a relatively large number of features were few (Sect.~\ref{rfit}), we also tested an
ANN algorithm, primarily to gauge just how small the sample could be before this failed to compete
with the machine learning models. Artificial neutral networks use a computer architecture based upon neurons in
a brain, where the artificial neuron is the functional unit of the network which
comprises several layers -- an input layer, weighting and biasing of the features, and several layers containing
non-linear activation functions, which transform the weighted input features. Each neuron may be connected from one to all
of the others in adjacent layers.  We used the {\sf TensorFlow}\footnote{https://www.tensorflow.org} platform,
with the testing of various hyperparameters giving the best results using a model similar to that in \citet{cmp21},
that is two {\em ReLu} layers and one {\em tanh} layer comprising $\gapp50$ to $\lapp200$ neurons each
(Fig.~\ref{fig:nn_architecture}). 
That is:
\begin{enumerate}
\item An  input layer with  one neuron for each training  feature, e.g. $\log_{10}\nu_{\text{TO}}$, $\log_{10}S_{\text{TO}}$, $\log_{10} (-X_2)$, $\log_{10} X_1$, $\log_{10} (-X_0)$, $\log_{10} S_{\text{1.4 GHz}}$, $\log_{10} S_{\text{5.0 GHz}}$, $\log_{10} S_{\text{8.7 GHz}}$ and $\alpha_{1.4~\text{GHz}}$, for the $n_{\text{int}}\geq1416$,  {\sc poly} sample.
\item  Three dense\footnote{Receiving input from all neurons in the preceding layer.} layers, each with 200 neurons, the first two using  $ReLu$ activation functions and  the third a $tanh$ function., 
\item One output layer for the target (the photometric redshift for the source).
\end{enumerate}

\subsection{Metrics}
\label{mam}

For each of the algorithms, we normalised the features and used a 80:20 training--validation split. We
quantified the quality of the redshift prediction according to:
\begin{enumerate}
\item The {\em regression coefficient}, $r$, of the least-squares linear fit between the predicted $(z_{\rm phot}$) and measured ($z_{\rm spec}$) redshifts. 
\item The {\em standard deviation} from the mean difference between the  photometric and spectroscopic redshifts, $\Delta z \equiv z_{\rm spec} - z_{\rm phot}$, 
\[
\sigma_{\Delta z} \equiv \sqrt{\frac{1}{N}\sum_{i=1}^N \Delta z^2},
\]
in addition to the {\em normalised standard deviation}, obtained from $\Delta z(\text{norm}) =  \Delta z /({z_{\rm spec} +1})$.
\item The {\em median absolute deviation} (MAD),
\[
\sigma_{\text{MAD}} \equiv 1.48 \times \text{median}\left|\Delta z\right|,
\]
in addition to the {\em normalised median absolute
  deviation} (NMAD),
\[
\sigma_{\text{NMAD}} \equiv 1.48 \times \text{median}\left|\Delta z(\text{norm})\right|.
\]
\item The outlier fraction (e.g. \citealt{cma+21,lzc+21}),
\[
\eta\equiv\frac{n_{|\Delta z(\text{norm})|> 0.15}}{n},
\]
that is, the number of sources with $|\Delta  z(\text{norm})|> 0.15$ as a fraction of the total number.
\end{enumerate}

\section{Results}
\subsection{Radio photometric redshifts}
\label{rpr}

As discussed in Sect.~\ref{rfit}, the number of sources available decreases rapidly as more radio-band 
features are added. Therefore, we test three different samples, ranging from the most features/least
sources to least features/most sources:

\begin{enumerate}
\item  Peaked spectrum sources only, where the turnover parameters are included as features.  
This  is split into two sub-samples:
\begin{enumerate}
\item The {\sc poly}  sample: Those for which a polynomial is fit, thus including $\log_{10}\nu_{\text{TO}}$ , $\log_{10}S_{\text{TO}}$, $\log_{10} (-X_2)$, $\log_{10} X_1$  and $\log_{10} (-X_0)$ as features. 
\item The {\sc snellen}  sample: The subset of these which could be fit by the function of \citet{ssd+98}, thus including $\log_{10} \nu_{\text{peak}}$, $\log_{10} S_{\text{peak}}$, $\alpha_{\text{thin}}$ and $\alpha_{\text{thick}}$ as features. 
\end{enumerate}
\item  The {\sc ohe}  sample: All of the sources, where the presence of a turnover in the SED was {\em one-hot encoded} as a feature. 
Thus, in addition to the flux densities
at different frequencies, the model contained two distinct sub-samples:
\begin{enumerate}
\item  Sources which exhibited a peak in the radio SED.
\item Sources where the SED exhibits no peak, which may be
\begin{enumerate}
  \item flat spectrum, with $|\alpha_{1.4~\text{GHz}}|\lapp0.5$, 
    \item steep spectrum,  with $\alpha_{1.4~\text{GHz}}\lapp-0.5$, or 
      \item inverted spectrum, with $\alpha_{1.4~\text{GHz}}\gapp0.5$,
\end{enumerate}
and where $\alpha_{1.4~\text{GHz}}$ may be correlated with the redshift (Sect.~\ref{rfit}).

\end{enumerate}
 \item  The {\sc flux}  sample: All of the sources, using only the flux densities at different frequencies as features. 
This is most analogous to the optical-band techniques, which utilise the source magnitudes.
\end{enumerate}
Since the 80:20 training--validation split applied to such small sample sizes can lead to  ``(un)lucky'' results, we ran
each algorithm 100 times, randomising the training and validation sources for each trial.

\subsubsection{Peaked spectrum sources}
\label{pss}


1403 of the sources with sufficient radio photometry exhibited a turnover which could be fit by the function of \citet{ssd+98}
and 2129 by a second order polynomial (Sect.~\ref{rfit}). However, referring to Table~\ref{intersect_all}, we see that the
inclusion of other features significantly decreases the sample size. We therefore used the features for which
$n_{\text{int}}\geq1416$, prioritised by the fit, starting with the polynomial (Table~\ref{intersect_all}).

When training and validating upon the NIR--optical--UV magnitudes (\citealt{cmp21} and references therein)  we achieve 
$r\gapp0.9$, $\sigma_{\Delta z (\text{norm})}\lapp0.3$ and  $\sigma_{\text{NMAD}}\lapp0.1$. Using this as a benchmark,  
from Table~\ref{ML_TO}  we see that all algorithms perform poorly
with little predictive power. 
\begin{table*}  
\begin{minipage}{170mm}
\centering
\caption{The mean values $\pm1\sigma$ of 100 trials of the machine learning algorithms, where the training and
validation sets are randomised for each trial. In addition to the metrics listed in Sect.~\ref{mam}, we quote the mean bias in 
$\Delta z$, $\mu_{\Delta z}$, and normalised bias,  $\mu_{\Delta z (\text{norm})}$.} 
\begin{tabular}{@{} l  c r c c r  c c  c @{}}
\hline
\smallskip
Regressor & & \multicolumn{3}{c}{Un-normalised} & \multicolumn{4}{c}{Normalised}\\
  &  $r$   & $\mu_{\Delta z}$ & $\sigma_{\Delta z}$ & $\sigma_{\text{MAD}}$ &  $\mu_{\Delta z (\text{norm})}$ &  $\sigma_{\Delta z (\text{norm})}$ & $\sigma_{\text{NMAD}}$ & $\eta$ \\
\hline
 \multicolumn{9}{c}{$n_{\text{int}}\geq1416$ of  the {\sc poly} sample  (284 validation sources)}\\
\hline
kNN &  $0.16\pm0.05$&   $-0.02\pm0.06$ & $0.92\pm0.04$  &  $0.92\pm0.06$ & $-0.16\pm0.03$ &$0.48\pm0.03$ &  $0.37\pm0.02$  & $67.6\pm2.5$\% \\ 
 SVR & $0.20\pm0.05$ &  $0.07\pm0.05$ & $0.90\pm0.04$  &  $0.93\pm0.06$ & $-0.11\pm0.03$ &  $0.45\pm0.02$ & $0.36\pm0.02$  &  $68.6\pm2.2$\%\\
DTR & $0.13\pm0.05$ &  $-0.01\pm0.06$ & $0.92\pm0.04$   & $0.96\pm0.06$ & $-0.15\pm0.03$ & $0.47\pm0.02$ & $0.38\pm0.02$  & $68.5\pm2.4$\% \\
ANN &   $0.17\pm0.04$ &  $0.00\pm0.28$ &  $0.96\pm0.02$  &  $0.89\pm0.06$&  $-0.14\pm0.08$ & $0.46\pm0.03$ & $0.36\pm0.07$ &  $67.8\pm2.3$\%\\
\hline
\multicolumn{9}{c}{$n_{\text{int}}\geq990$ of the {\sc snellen} sample (198  validation sources)}\\
\hline
kNN &  $0.14\pm0.07$ &   $-0.02\pm0.06$ & $0.93\pm0.05$ &  $0.95\pm0.07$  &$-0.16\pm0.03$ &  $0.48\pm0.03$  &  $0.38\pm0.03$ & $70.0\pm2.8$\% \\ 
SVR &  $0.14\pm0.06$  &  $0.06\pm0.07$ & $0.92\pm0.05$  &  $0.94\pm0.07$ & $-0.12\pm0.03$ & $0.46\pm0.03$ & $0.38\pm0.03$  &  $71.3\pm2.7$\%\\
DTR & $0.06\pm0.06$ &  $0.01\pm0.08$ &  $0.94\pm0.04$ &  $0.99\pm0.07$ &  $-0.15\pm0.04$ &  $0.48\pm0.03$ &  $0.39\pm0.03$ &  $69.7\pm3.0$\% \\ 
ANN &  $0.13\pm0.04$ &  $0.17\pm0.16$ & $0.97\pm0.02$  &  $0.94\pm0.05$  &  $-0.07\pm0.07$ & $0.46\pm0.03$  & $0.38\pm0.02$ & $70.7\pm2.8$\% \\
\hline
\multicolumn{9}{c}{$n_{\text{int}}\geq1001$ of the {\sc ohe} sample (201 validation sources)}\\
\hline 
kNN &  $0.34\pm0.06$ &   $-0.04\pm0.07$ & $0.83\pm0.04$  & $0.83\pm0.06$   & $-0.15\pm0.04$ & $0.44\pm0.03$ &  $0.36\pm0.03$  & $67.7\pm3.0$\%\\
SVR & $0.35\pm0.05$ &  $0.05\pm0.07$ & $0.82\pm0.04$   &  $0.79\pm0.07$  & $-0.11\pm0.04$ & $0.43\pm0.03$  & $0.34\pm0.03$  & $65.7\pm3.1$\%\\
DTR & $0.30\pm0.06$ &  $-0.00\pm0.07$ & $0.85\pm0.04$ & $0.87\pm0.07$ & $-0.14\pm0.04$ & $0.44\pm0.03$ &  $0.37\pm0.04$ & $68.2\pm3.1$\%\\
ANN & $0.30\pm0.03$  &  $-0.08\pm0.12$ & $0.93\pm0.01$  &  $0.91\pm0.05$ &  $-0.20\pm0.06$  & $0.49\pm0.03$ & $0.40\pm0.03$  &  $70.5\pm2.6$\%\\
\hline
\multicolumn{9}{c}{$n_{\text{int}}\geq351$ of the {\sc ohe} sample (71 validation sources)}\\
\hline 
kNN & $0.41\pm0.09$ &   $ -0.01\pm 0.10 $ &  $0.77\pm0.08$ & $0.72\pm 0.09$ &  $-0.13\pm 0.05$ & $0.40\pm0.05$ & $0.38\pm  0.05$ & $70.5 \pm5.0$\% \\
SVR  & $0.46\pm0.09$ & $0.09\pm0.10$&  $0.77\pm0.08$  & $0.70\pm0.09$ &  $ -0.06\pm0.05$& $ 0.40\pm 0.05$ & $0.34\pm 0.05$ & $ 65.5\pm5.2$\% \\
DTR & $0.31\pm0.11$ & $ 0.01\pm 0.12$ & $0.84\pm 0.08$ &  $0.75\pm 0.09$ &  $-0.12\pm 0.06$ & $ 0.43\pm0.06$ & $0.37\pm 0.04$ &  $70.0\pm5.0$\%\\
ANN &  $0.39\pm 0.03$ &   $-0.13\pm0.16$ & $0.80\pm0.02$ & $0.79\pm 0.08$ &$-0.19\pm0.08$ &   $0.43\pm 0.03$ &   $0.36 \pm0.05$ & $67.2\pm4.4$\% \\  
\hline
\multicolumn{9}{c}{$n_{\text{int}}\geq2037$ of the  {\sc flux} sample (408 validation sources)}\\
\hline  
 kNN & $0.29\pm0.49$ &   $ -0.03\pm 0.04 $ &  $0.78\pm0.03$ & $0.80\pm0.04$ &   $-0.14\pm0.02$ & $0.41\pm0.01$ &  $0.37\pm0.02$ & $68.6\pm1.8$\%\\
SVR & $0.30\pm0.04$ & $0.10\pm0.04$ & $0.78\pm0.03$ &  $0.76\pm0.04$  & $-0.08\pm0.02$  &  $0.39\pm0.02$  & $0.36\pm0.02$ & $68.1\pm2.1$\%\\
DTR &  $0.23\pm0.04$  &  $0.00\pm0.04$ & $0.80\pm0.03$  & $0.81\pm0.04$   & $-0.13\pm0.02$ & $0.43\pm0.02$   & $0.37\pm0.02$ & $69.1\pm2.0$\%\\
ANN & $0.28\pm0.04$ & $0.00\pm0.10$ & $0.78\pm0.03 $ & $0.80\pm0.05$ &  $-0.13\pm\pm0.05$ &  $0.42\pm0.02$ &   $0.36\pm 0.02$ & $67.9\pm2.4$\% \\ 
 \hline
\multicolumn{9}{c}{$n_{\text{int}}\geq716$ of the {\sc flux}  sample (144 validation sources)}\\
\hline 
kNN & $0.36\pm0.06$   & $-0.03\pm0.07$ &$0.73\pm0.05$   & $0.73\pm0.06$   & $-0.13\pm0.04$ & $0.39\pm0.02$  & $0.38\pm0.03$  & $69.4\pm3.6$\%\\
SVR  & $0.38\pm0.06$  & $0.09\pm0.06$ & $0.72\pm0.04$  &  $0.65\pm0.05$   & $-0.06\pm0.03$ & $0.36\pm0.02$  &  $0.34\pm0.03$   & $65.9\pm3.9$\%\\
DTR &  $0.31\pm0.07$  & $0.01\pm0.07$ & $9.76\pm0.06$  &  $0.73\pm0.07$   & $-0.11\pm0.04$ & $0.40\pm0.03$   & $0.37\pm0.03$   & $67.4\pm4.0$\%\\
ANN & $0.33\pm0.08$  &  $0.02\pm0.20$ & $0.75\pm0.05$  & $0.73\pm0.09$  &  $-0.10\pm0.10$ & $0.40\pm0.04$  & $0.37\pm0.04$ &  $69.9\pm4.4$\% \\
 \hline
\multicolumn{9}{c}{$n_{\text{int}}\geq549$ of the {\sc flux}  sample (110 validation sources)}\\
 \hline 
kNN &  $0.36\pm0.08$  & $-0.03\pm0.08$ & $0.66\pm0.04$   & $0.67\pm0.06$  & $-0.12\pm0.05$ & $0.35\pm0.03$  & $0.36\pm0.03$  & $68.2\pm4.0$\% \\ 
SVR & $0.37\pm0.06$  & $0.09\pm0.07$ & $0.66\pm0.07$ &  $0.60\pm0.06$ &  $-0.05\pm0.04$ & $0.33\pm0.02$  & $0.33\pm0.33$  & $63.6\pm4.3$\%\\ 
DTR &  $0.30\pm0.08$  & $0.00\pm0.08$ & $0.70\pm0.04$  & $0.68\pm0.07$  & $-0.10\pm0.04$ & $0.37\pm0.03$  & $0.36\pm0.04$  &  $67.3\pm4.4$\%\\
ANN &  $0.30\pm0.09$  &  $0.01\pm0.14$ & $0.70\pm0.05$  &  $0.70\pm0.10$  & $-0.10\pm0.08$ & $0.38\pm0.04$  & $0.37\pm0.05$  &  $68.2\pm5.1$\% \\
\hline
\multicolumn{9}{c}{$n_{\text{int}}\geq329$ of the  {\sc flux} sample  (66 validation sources)}\\
\hline
 kNN & $0.26\pm0.11$  & $-0.05\pm0.09$ & $0.63\pm0.06$   & $0.67\pm0.08$   & $-0.13\pm0.05$ & $0.34\pm0.03$   & $0.37\pm0.04$  & $67.8\pm5.6$\%\\
SVR &  $0.27\pm0.09$  & $0.09\pm0.10$ & $0.64\pm0.06$   & $0.58\pm0.08$  &$-0.05\pm0.05$ &  $0.32\pm0.03$  &  $0.31\pm0.05$   & $62.6\pm5.3$\%\\
DTR & $0.27\pm0.11$  &  $0.00\pm0.09$ & $0.65\pm0.06$  &  $0.65\pm0.09$   & $-0.10\pm0.05$ & $0.35\pm0.03$   & $0.35\pm0.04$   & $64.8\pm5.2$\%\\
ANN & $0.19\pm0.12$  &  $-0.001\pm0.17$ & $0.69\pm0.06$  &  $0.72\pm0.10$ &  $-0.10\pm0.09$ & $0.39\pm0.04$  & $0.39\pm0.05$ & $ 69.4\pm5.5$\%\\
\hline
\end{tabular}
\label{ML_TO}  
\end{minipage}
\end{table*} 
Adding the next populous feature,  $\log_{10} S_{\text{400 MHz}}$, giving $n_{\text{int}}\geq559$ and a 
validation sample size of 112, gave no improvement in the predictions.
Repeating the procedure for the SED fit of \citeauthor{ssd+98}, gives
the sample sizes listed in Table~\ref{intersect_all} and
again, poor results (Table~\ref{ML_TO}).

\subsubsection{All radio sources}

\begin{figure}
\centering \includegraphics[angle=-90,scale=0.52]{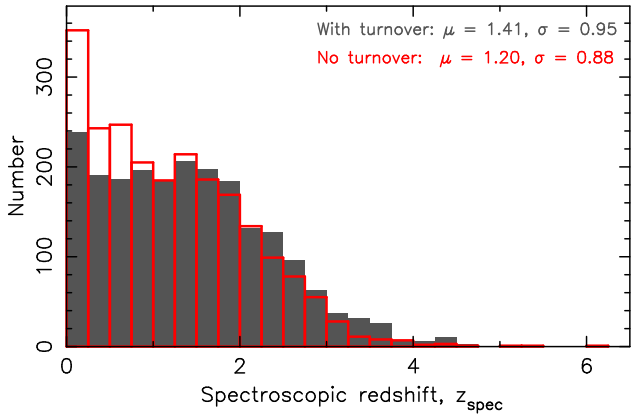}
\caption{The redshift distributions of the radio sources with and without turnovers in their SEDs.
A Kolmogorov-Smirnov test gives a probability of $7.86\times10^{-9}$ of the sources being drawn from the same sample.}
\label{TO_OHE}
\end{figure}

Testing all of the radio sources, while wishing to retain the presence of a turnover in the SED as a feature (Fig.~\ref{TO_OHE}),
we ``one-hot encoded''  this feature as a binary value.
From the remaining features we again shortlisted those in Table~\ref{intersect_all}
with $n_{\text{int}}\geq1001$. Testing this, we see that all algorithms gave a significant
improvement in the regression coefficient over the previous models,
although the other metrics show that these still have little predictive power.

\begin{figure} 
\centering \includegraphics[angle=-90,scale=0.5]{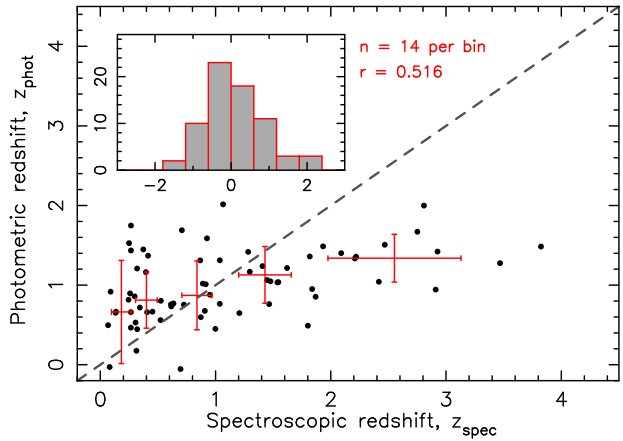}
\caption{The photometric versus the spectroscopic redshift for the 71 validation sources for  a single run of the 
SVR algorithm on the {\sc ohe} $n_{\text{int}}\geq351$ sample. 
The binning is for an equal number of sources in each bin with the error bars showing $\pm1\sigma$. 
The broken line shows $z_{\text{phot}} = z_{\text{spec}}$ and the  inset the distribution of $\Delta z$.}
\label{pred_OHE}
\end{figure}

By removing the need for the SEDs to be fit by either a polynomial or the \citeauthor{ssd+98} function significantly
increases the sample sizes, by utilising only the fluxes in the model (Table~\ref{intersect_all}).  This is analogous to
the NIR--optical--UV model (Sect.~\ref{stv}) and, like the one-hot encoded sample, this outperforms both the 
{\sc poly} and {\sc snellen} samples. From Table~\ref{ML_TO}, we see that the samples with $n_{\text{int}}\geq716$  and $n_{\text{int}}\geq549$ give the best results, with no difference between the two models within the $1\sigma$ uncertainties, 
despite the significant drop in sample sizes from $n_{\text{int}}\geq2037$. This suggests that the low frequency flux
densities ($S_{\text{70 MHz}}$ \&  $S_{\text{150 MHz}}$) may be important features in the prediction of photometric redshifts.
We therefore, also tested the {\sc ohe} sample including the next numerous feature, $S_{\text{150 MHz}}$, giving $n_{\text{int}}\geq351$.
Although the numbers are small, just 71 validation sources, we see further improvement in the statistics.
However, within the larger $1\sigma$ uncertainties the results are consistent with the $n_{\text{int}}\geq1001$ {\sc ohe} 
results for all of the machine learning algorithms. We would therefore err towards the $r\approx0.39$ given by the
ANN trials as being representative. There is also the possibility that the relatively good statistics are due to the much smaller test sample.
In any case, 
these  still fall far short of being able to provide photometric redshift predictions (Fig.~\ref{pred_OHE}).

\begin{figure} 
\centering \includegraphics[angle=-90,scale=0.53]{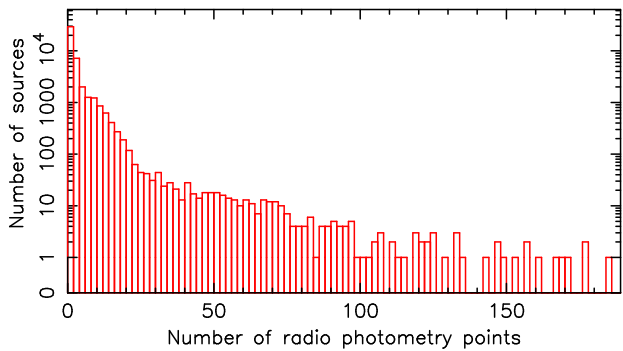}
\caption{The distribution of the number of radio photometry measurements.}
\label{n_radio}
\end{figure}

Lastly, the  fact that the {\sc ohe} and {\sc flux} samples provide better models than the {\sc poly} and {\sc snellen} samples
suggests there may be something wrong with the fitting. This is a possibility since the fits were performed on
heterogeneously sampled data, with the requirement of at least three photometry measurements (9833 sources) for the 
polynomial fit and at least five for the \citeauthor{ssd+98} fit (5604 sources, of which not all could be fit).
This can lead to clear under-sampling of some SEDs, possibly giving unrepresentative fits, although raising this minimum requirement drastically cut the sample sizes still further (Fig.~\ref{n_radio}).
Whether or not a fit is present does, however, seem to provide a useful feature (the {\sc ohe} sample) and so with
better sampling of the SEDs, giving more accurate measurements of the fit parameters, these may yet
prove to be useful features.

\subsubsection{Effectiveness of the machine learning}

Although the results are indicative of the radio data being  ineffective in providing a redshift prediction model, we
do see some quite strong variation across the models (e.g. the regression coefficient ranging from $r\approx0.06$ to
0.46, Table~\ref{ML_TO}). While the numbers are small and it is not known if the most numerous features are the most optimal
(Table~\ref{intersect_all}), we can test if the algorithms are at least partially effective in predicting
photometric redshifts.

\begin{figure} 
\centering \includegraphics[angle=-90,scale=0.5]{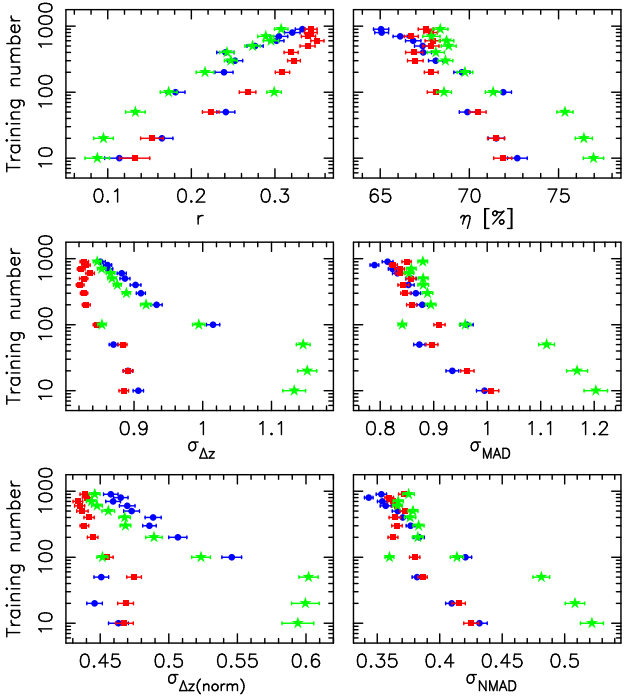}
\caption{The results for the kNN (red), SVR (blue) and DTR (green) one-hot encoded models ($n_{\text{int}}\geq1001$) based
on different training sample sizes. Note that for clarity the error bars show the standard error about the mean.}
\label{frac_test}
\end{figure}

For this, we select the $n_{\text{int}}\geq1001$  {\sc ohe} sample, 
since this yields some of the highest values of $r$ for all algorithms, while
providing a relatively large sample.  For each algorithm we train on progressively fewer sources and validate on
another 100. That is, training on 900, 800, ..., 100, 50, 20,10  sources, randomly selected for each of the 100 trials.
From the results (Fig.~\ref{frac_test}),
we see that all models  are  indicative of the algorithms being effective, with an increase in
$r$ with the training size along with a decrease in the spread in $\Delta z$. Note, however, that the figure shows the
standard errors,  since the $\pm1\sigma$ ranges exceed the abscissa ranges. For instance, for the SVR model
$r = 0.11\pm0.18$ training on 10 sources and
$0.33\pm0.08$ training on 900. So while the mean values appear to improve with the training sample size, the
uncertainties are too large to state this definitively.

\subsection{NIR--optical--UV photometric redshifts}

\subsubsection{Training and validation}
\label{stv}

\begin{figure} 
\centering \includegraphics[angle=-90,scale=0.5]{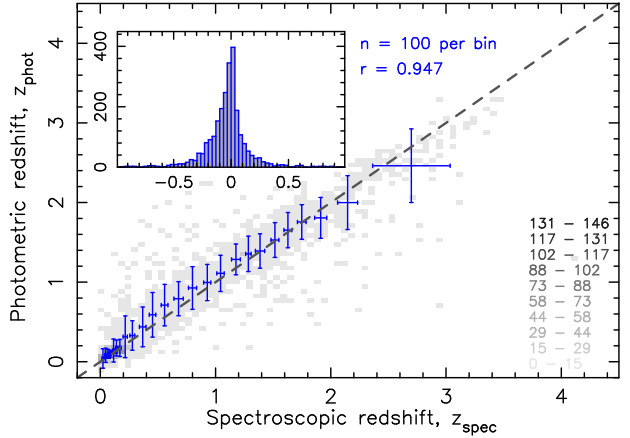}
\caption{An instance of the photometric redshift predictions from the NIR--optical--UV photometry versus the spectroscopic redshift 
for the 2501 Milliquas validation sources,
trained upon the other 10\,002  Milliquas radio sources. The grey-scale shows the source 
distribution with the key to the bottom right showing the number within each pixel.
Here $\mu_{\Delta z}= -0.03$, $\sigma_{\Delta z} = 0.24$, 
$\sigma_{\text{MAD}}=0.11$, $\mu_{\Delta z (\text{norm})}= -0.03$, $\sigma_{\Delta z(\text{norm})} = 0.13$,  $\sigma_{\text{NMAD}}=0.06$ and $\eta = 11$\%.}
\label{pred_TF-mags}
\end{figure}

We also tested the predictive power of the machine learning models, using the NIR--optical--UV photometry, which has
proven successful in training and validation upon sources from a single dataset (i.e. the SDSS, e.g. \citealt{bmh+12,bcd+13}),
as well as training on one dataset and validating upon another (i.e. the SDSS on radio-selected quasars from external catalogues,
\citealt{cur20,cmp21}). Of the Milliquas radio sources, there were 12\,503 which had all of the $W2, W1, z,i,r,g,u,NUV,FUV$ magnitudes,
which we used directly as features.\footnote{The usual practice is to use the $u- g$, $g - r$, $r - i$ \& $ i - z$
  colours, although \citet{cur22} finds the magnitudes to perform equally as well.}
\begin{table*}  
\centering
\caption{The mean values $\pm1\sigma$ of 100 trials of the machine learning algorithms using the NIR--optical--UV magnitudes (2501 validation sources).}
\begin{tabular}{@{} l  c r c c r  c c  c @{}}
\hline
\smallskip   
   & & \multicolumn{3}{c}{Un-normalised} & \multicolumn{4}{c}{Normalised}\\
  &  $r$   & $\mu_{\Delta z}$ & $\sigma_{\Delta z}$ & $\sigma_{\text{MAD}}$ &  $\mu_{\Delta z (\text{norm})}$ &  $\sigma_{\Delta z (\text{norm})}$ & $\sigma_{\text{NMAD}}$ & $\eta$ \\
\hline
kNN &  $0.911\pm0.006$ &   $0.019\pm0.008$ &  $0.308\pm0.011$  & $0.141\pm0.005$   & $-0.013\pm0.004$ & $0.143\pm0.004$  &  $0.077\pm0.002$  & $17.7\pm0.7$ \% \\
SVR & $0.913\pm0.005$ &  $0.016\pm0.006$ & $0.283\pm0.009$  &  $0.149\pm0.004$ & $-0.008\pm0.003$ & $0.142\pm0.043$  &$0.089\pm0.002$  & $17.0\pm0.8$\%\\
DTR & $0.858\pm0.009$ &  $0.002\pm0.009$ & $0.387\pm0.013$ & $0.170\pm0.007$ & $-0.022\pm0.004$ & $0.204\pm0.011$ & $0.091\pm0.003$  & $22.6\pm0.8$\%\\
ANN &  $0.944\pm0.003$ & $-0.005\pm0.030$ & $0.249\pm0.006$   & $0.115\pm0.012$ & $-0.016\pm0.018$ & $0.124\pm0.006$  &$0.066\pm0.009$   & $11.6\pm1.4$\%\\
\hline
\end{tabular}
\label{ML_mag}  
\end{table*} 
From Table~\ref{ML_mag}, we see that all models are successful, with the ANN (Fig.~\ref{pred_TF-mags}) being 
comparable with the SDSS model of \citet{cmp21}.

That is, using the NIR--optical--UV features appears to perform significantly better than the radio selected features.
This, however, is based upon a much larger sample for which the training may be more comprehensive.  In order
to test whether a larger training sample was the reason for the NIR--optical--UV model's superiority, we selected
a similar number of sources (1000) at random from those with the complete $W2, W1, z,i,r,g,u,NUV,FUV$ photometry.
Again, using the 80:20 training--validation split, we would train on 800 of the sources and validate on the remaining 200,
before selecting another 1000 sources at random and repeating until 100 trials were complete.
\begin{table*}  
\centering
\begin{minipage}{155mm}
\caption{The results of the machine learning algorithms for a sample of 1000 unique sources selected at
random from the Milliquas sample. We run each algorithm 100 times, randomising the 1000 sources
each time and so quote the mean values, with $\pm1\sigma$.} 
\begin{tabular}{@{} l  c r c c r  c c  c @{}}
\hline
\smallskip
   & & \multicolumn{3}{c}{Un-normalised} & \multicolumn{4}{c}{Normalised}\\
  &  $r$   & $\mu_{\Delta z}$ & $\sigma_{\Delta z}$ & $\sigma_{\text{MAD}}$ &  $\mu_{\Delta z (\text{norm})}$ &  $\sigma_{\Delta z (\text{norm})}$ & $\sigma_{\text{NMAD}}$ & $\eta$ \\
\hline
kNN & $0.87\pm0.03$ &  $0.03\pm0.03$ &  $0.37\pm0.05$ &  $0.22\pm0.02$ &  $-0.02\pm0.01$ &  $0.17\pm0.02$ &  $0.12\pm0.01$ &  $29.0\pm3.1$\%\\
SVR &  $0.89\pm0.03$ &  $0.01\pm0.03$ &  $0.33\pm0.05$ &  $0.20\pm0.02$ &  $-0.02\pm0.01$ &  $0.17\pm0.02$ &  $0.12\pm0.01$ &  $25.3\pm3.1$\%\\ 
DTR & $0.78\pm0.04$ &  $0.01\pm0.04$ &  $0.46\pm0.04$ &  $0.29\pm0.04$ &  $-0.04\pm0.02$ &  $0.23\pm0.03$ &  $0.16\pm0.02$ &  $38.3\pm3.7$\% \\
ANN & $0.90\pm0.03$ &  $0.00\pm0.09$ &  $0.33\pm0.04$ &  $0.20\pm0.03$ &  $-0.02\pm0.04$ &  $0.16\pm0.03$ &  $0.12\pm0.02$ &  $24.1\pm5.3$\%\\ 
\hline
\end{tabular}
\label{ML_mag-1000}  
\end{minipage}
\end{table*} 
The results are summarised in Table~\ref{ML_mag-1000}, from which we see the NIR--optical--UV model 
still provides accurate photometric redshifts, thus confirming that
the $W2, W1, z,i,r,g,u,NUV,FUV$ photometry  provide superior features to those in the radio band.

\subsubsection{SDSS training of Milliquas sources}
\label{stms}

\begin{figure*}
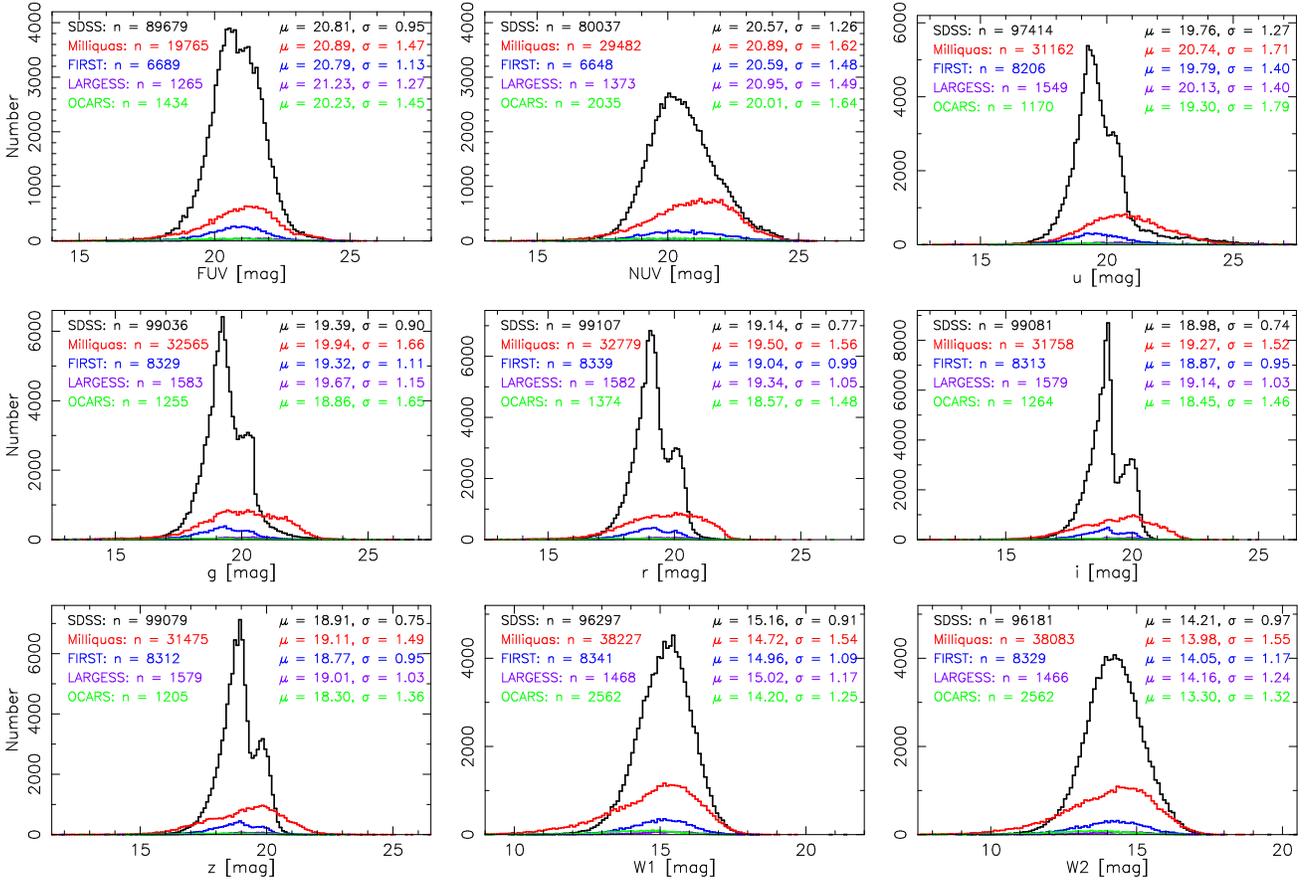

\centering \includegraphics[angle=-90,scale=0.35]{FUV-histo_140.eps}
\centering \includegraphics[angle=-90,scale=0.35]{NUV-histo_145.eps}
\centering \includegraphics[angle=-90,scale=0.35]{u-histo_150.eps}
\centering \includegraphics[angle=-90,scale=0.35]{g-histo_150.eps}
\centering \includegraphics[angle=-90,scale=0.35]{r-histo_150.eps}
\centering \includegraphics[angle=-90,scale=0.35]{i-histo_150.eps}
\centering \includegraphics[angle=-90,scale=0.35]{z-histo_150.eps}
\centering \includegraphics[angle=-90,scale=0.35]{W1-histo_130.eps}
\centering \includegraphics[angle=-90,scale=0.35]{W2-histo_130.eps}
\caption{The distribution of each of the magnitudes for the sample compared to the SDSS and our three radio selected
catalogues (see Sect.~\ref{rss}).}
\label{SDSS_histos}
\end{figure*}

We also tested the prediction of the redshifts of the Milliquas sources, based upon our SDSS model. That is, 
using the 71\,267 of the 100\,337 SDSS QSOs with the full 
NIR--optical--UV photometry \citep{cmp21} to train a model which we  then validate on the Milliquas radio sources.
This frees up the Milliquas data from training, giving  photometric
redshifts for all 12\,503 sources with the full magnitude complement and, from 
\begin{figure} 
\centering \includegraphics[angle=-90,scale=0.5]{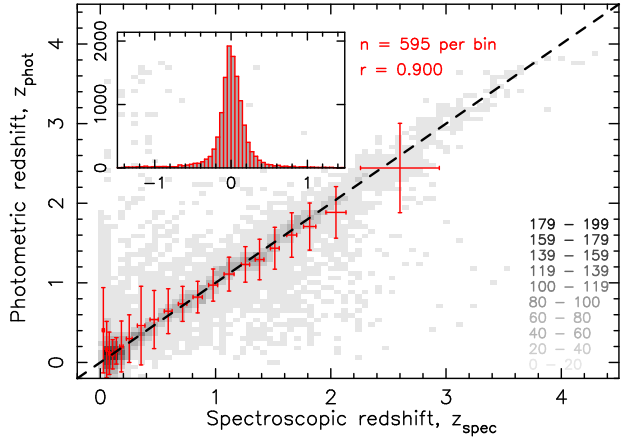}
\caption{Validation of  Milliquas redshifts predicted  from an NIR--optical--UV photometry model trained on the SDSS sample of \citet{cmp21}. 
Here $n = 12\,503$, $\mu_{\Delta z}= 0.00$, $\sigma_{\Delta z} = 0.33$,  $\sigma_{\text{MAD}}=0.14$, $\mu_{\Delta z (\text{norm})}= -0.02$, $\sigma_{\Delta z(\text{norm})} = 0.23$, $\sigma_{\text{NMAD}}=0.08$ and $\eta=19$\%.}
\label{pred_X-SDSS}
\end{figure} 
Fig.~\ref{pred_X-SDSS}, we see that the SDSS provides a good model with metrics close
to that of other external datasets validated upon the SDSS model (\citealt{cmp21}, see also Sect.~\ref{rss}). This is despite
quite different distributions in the magnitudes between all of the tested data (Fig.~\ref{SDSS_histos}).

\begin{figure}
\centering \includegraphics[angle=-90,scale=0.48]{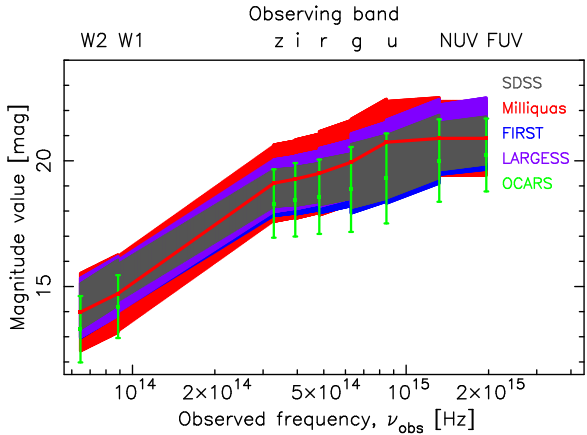}
\caption{The ``magnitude profiles'' of the various databases, where the coloured bands show the mean magnitude $\pm1\sigma$ (Fig.~\ref{SDSS_histos}). For clarity,  OCARS (see Sect.~\ref{rss}) is shown as error bars and the mean magnitudes from the Milliquas sources have also been over-plotted as a line.}
\label{mags_sum}
\end{figure}

Compared to training the model on the Milliquas sample (Fig.~\ref{pred_TF-mags}), 
we see the majority of the scatter to occur at $z\sim0$.   We have previously noted the 
lower redshifts to be problematic in our SDSS sample, probably due to the relatively unreliable FUV fluxes
which are important in  utilising  the $\lambda= 1216$~\AA\ Lyman-break in the model 
\citep{cmp21}.  From the ``magnitude profiles'' of the datasets (Fig.~\ref{mags_sum}),
we confirm that the Milliquas magnitudes differ from those of the SDSS, with the former being flatter
across $NUV$ and $FUV$.

\subsubsection{Milliquas training of radio selected sources}
\label{rss}

A major issue with training a model using SDSS photometry is that these are restricted to the northern sky only, whereas
the SKA and its pathfinders will survey  the southern sky. 
In \citet{cmp21} we suggested that {\em SkyMapper} \citep{wol+18}\footnote{https://skymapper.anu.edu.au}, 
which will survey the southern sky in the $u, v, g, r, i, z$ bands, could provide the photometry for the
SKA sources.  Until the SkyMapper data become generally available, we can test if alternative datasets to the 
SDSS are viable using the Milliquas data. Although these are  also predominately  located  in the northern sky (see Sect.~\ref{imp}), 
their magnitude distributions are quite distinct from those of the SDSS (Fig.~\ref{SDSS_histos}).

\begin{figure}
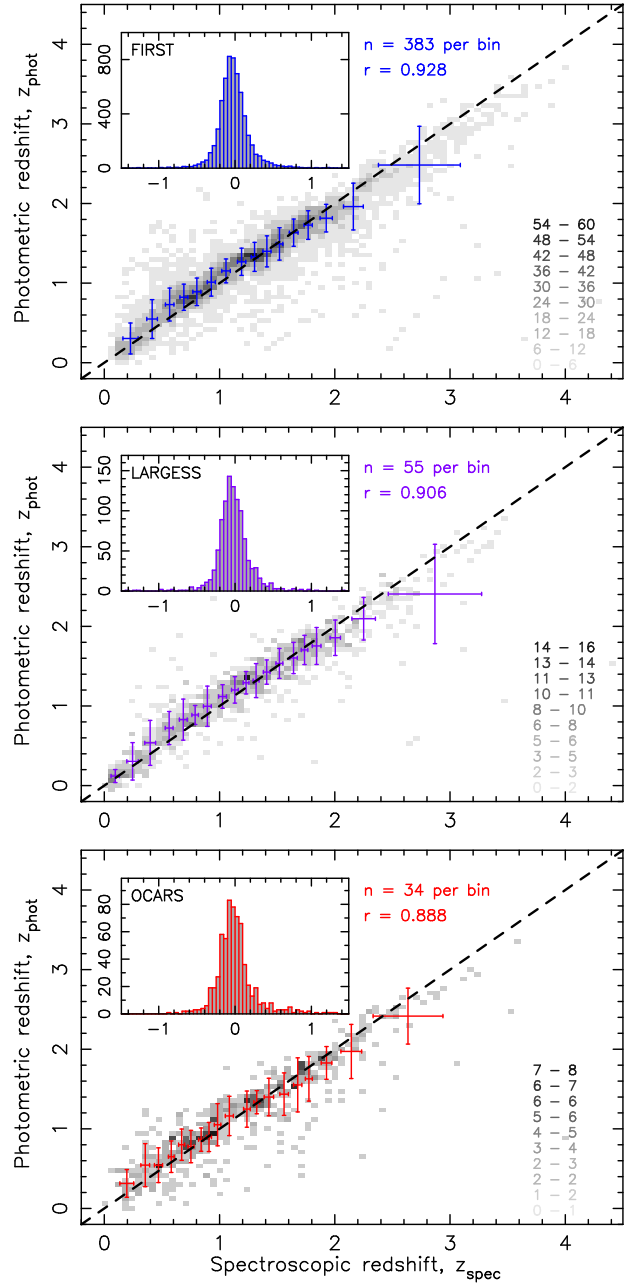
 
\centering \includegraphics[angle=-90,scale=0.5]{pred_X-FIRST_383.eps}
\centering \includegraphics[angle=-90,scale=0.5]{pred_X-LARGESS_55.eps}
\centering \includegraphics[angle=-90,scale=0.5]{pred_X-OCARS_34.eps}
\caption{Validation of the redshifts from the three external radio-selected databases from a model trained on the Milliquas sources. Top: FIRST -- $n = 6129$, $\mu_{\Delta z}= -0.01$, $\sigma_{\Delta z} = 0.25$,  $\sigma_{\text{MAD}}=0.16$,  $\mu_{\Delta z (\text{norm})}= -0.02$, $\sigma_{\Delta z(\text{norm})} = 0.11$,  $\sigma_{\text{NMAD}}=0.07$ and $\eta = 11\%$. 
Middle: LARGESS -- $n = 1046$, $\sigma_{\Delta z} = 0.30$,  $\mu_{\Delta z}= 0.0$, $\sigma_{\text{MAD}}=0.16$, $\mu_{\Delta z (\text{norm})}= -0.02$, $\sigma_{\Delta z(\text{norm})} = 0.12$, $\sigma_{\text{NMAD}}=0.08$ and $\eta = 10$\%. Bottom: OCARS -- $n=649$, $\mu_{\Delta z}= 0.02$, $\sigma_{\Delta z} = 0.31$,  $\sigma_{\text{MAD}}=0.16$, $\mu_{\Delta z (\text{norm})}= -0.01$, $\sigma_{\Delta z(\text{norm})} = 0.11$,  $\sigma_{\text{NMAD}}=0.08$ and $\eta = 18$\%.}
\label{pred_X}
\end{figure}

As discussed in \citet{cm19}, large catalogues of radio sources with spectroscopic redshifts are rare and so we use 
the same samples as previously \citep{cmp21}:
\begin{enumerate}

\item  The {\em Faint Images of the Radio Sky at Twenty-Centimeters} (FIRST, \citealt{bwh95,wbhg97}) sample.
Of the 18\,273 sources with redshifts from the SDSS DR14 QSOs \citep{ppa+18} 9016 are classified in NED as QSOs.

\item The {\em Large Area Radio Galaxy Evolution Spectroscopic Survey} (LARGESS). Of the 10\,685 sources with 
  optical redshifts \citep{csc+17}\footnote{Those with redshift reliability flag $q\geq3$, where 
$q=3$  designates ``a reasonably confident redshift'',  and the maximum $q=5$ designates an ``extremely reliable redshift from a good-quality spectrum''.} 1608  are classified in NED as QSOs.

\item The {\em Optical Characteristics of Astrometric Radio Sources} (OCARS) catalogue of {\em Very Long Baseline
  Interferometry} (VLBI) astrometry sources \citep{mab+09,mal18}. Of the 7800 sources, 3151
are classified in NED as QSOs.
\end{enumerate}
In Fig.~\ref{pred_X}, we show the predicted redshifts from the validation of the sources in the above catalogues
trained upon the Milliquas radio sources.
The results are largely indistinguishable from the SDSS training (Table~\ref{SDSS_Miiliqua}),
even though the SDSS sample is significantly larger and we had previously used the colours, rather than magnitudes, as features.
\begin{table*}
\centering
\begin{minipage}{140mm}
\caption{Comparison of  the radio catalogue predictions from the SDSS (71\,267 QSOs, 
\citealt{cmp21}) and Milliquas (44\,119 quasars) magnitudes.}
\begin{tabular}{@{}l  cc  cc cc cc cc   @{}}
\hline
\smallskip
Sample  & \multicolumn{2}{c}{$r$} & \multicolumn{2}{c}{$\sigma_{\Delta z}$} & \multicolumn{2}{c}{$\sigma_{\text{MAD}}$} & \multicolumn{2}{c}{$\sigma_{\Delta z (\text{norm})}$} & \multicolumn{2}{c}{$\sigma_{\text{NMAD}}$}\\
             & SDSS & Milliquas & SDSS & Milliquas &  SDSS & Milliquas & SDSS & Milliquas & SDSS & Milliquas\\
\hline
FIRST & 0.93 &  0.93   & 0.245 & 0.248& 0.102 & 0.158& 0.110 & 0.111& 0.048 & 0.072\\
LARGESS & 0.91 & 0.91 & 0.297 & 0.392 & 0.123 & 0.163 & 0.123 & 0.120 & 0.057 & 0.075\\
OCARS & 0.85 & 0.89 & 0.371 & 0.307 & 0.132 & 0.164 & 0.170 & 0.131 & 0.063 & 0.076\\
\hline
\end{tabular}
\label{SDSS_Miiliqua}  
\end{minipage}
\end{table*}

\subsection{Imputation of missing data} 
\label{imp}

As mentioned in Sect.~\ref{stv}, only 12\,503 of the 44\,119 Milliquas sources have the full magnitude complement.
Like the OCARS sources, much of this can be attributed to Milliquas covering the whole sky, whereas the SDSS
\begin{figure*}
\includegraphics[angle=0,scale=0.7]{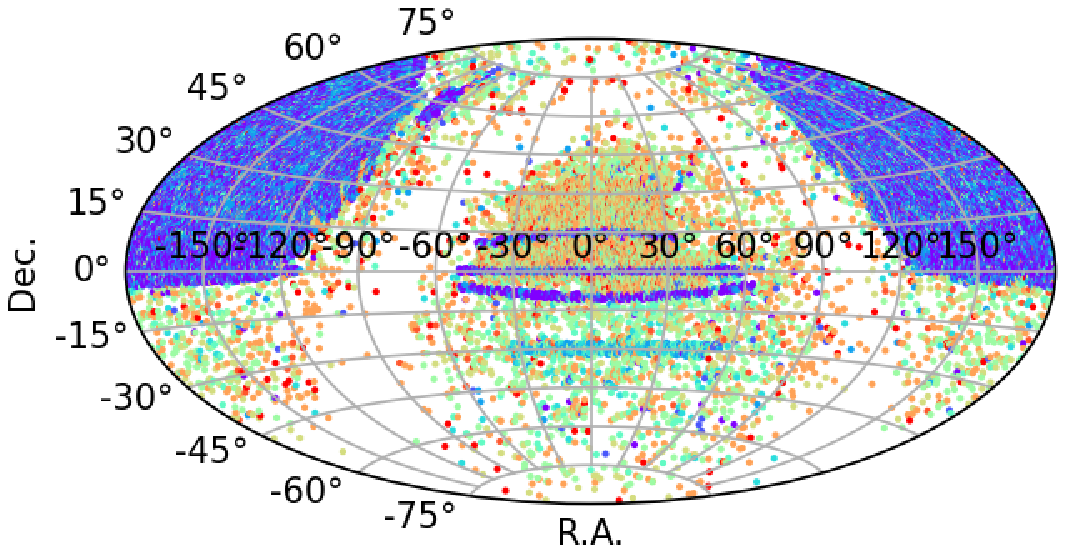}
\includegraphics[angle=0,scale=0.7]{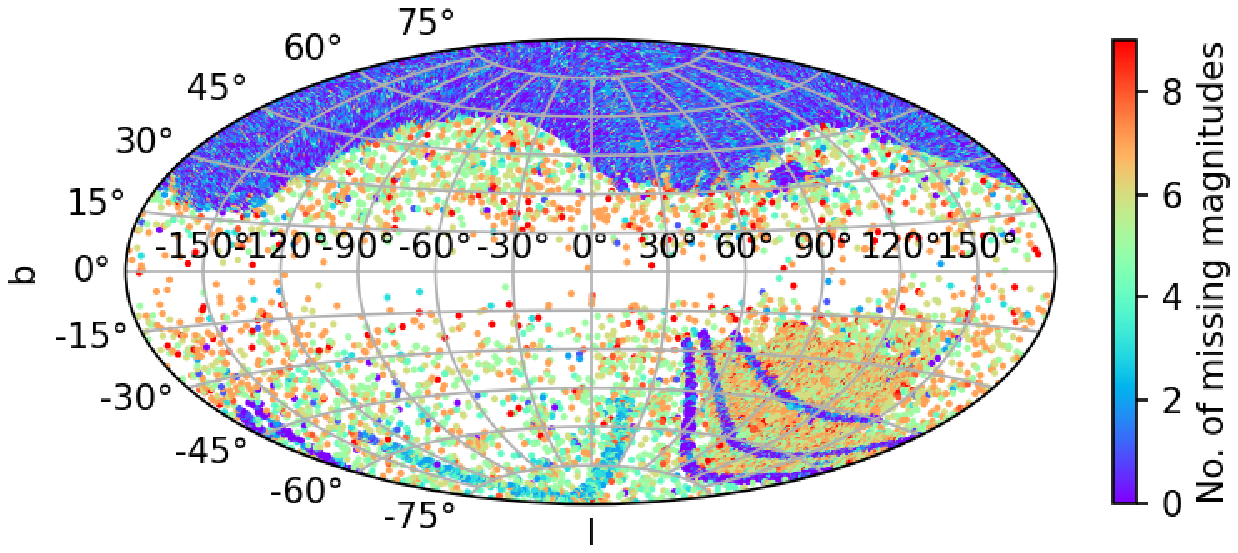}
\caption{The sky distribution of the Milliquas radio sources in equatorial (left) and Galactic (right) coordinates, colour coded by the number of missing
magnitudes per source.}
\label{sky}
\end{figure*}
is limited to the northern sky (Fig.~\ref{sky}, left). However, we also see that a significant amount of 
 GALEX photometry is missing (Table~\ref{missing}).
\setlength{\tabcolsep}{0.4em}
\begin{table}
\centering
\caption{The number of missing magnitude measurements in each of the bands for the Milliquas sources.}
\begin{tabular}{@{}c c c c c c c c c c @{}}
\hline
\smallskip 
 $FUV$ &    $NUV $  & $u$ &  $g$ &    $r$  &     $i$ &   $z$ &    $W1$ &    $W2$ &    \\
24\,365 & 14\,646 & 12\,968 & 11\,565 & 11\,351 & 12\,372 & 12\,655 & 5903 & 6047 \\
\hline
\end{tabular}
\label{missing}  
\end{table} 
Given the similarity between the distribution of  missing magnitudes in Galactic coordinates (Fig.~\ref{sky}, right) 
and the $NUV \cap FUV \cap$\,SDSS distribution \citep{bcs14}, we, again, attribute this to the sky coverage.

In combination with the missing WISE magnitudes, this means that we can only predict redshifts for 
28\% of the sample. We therefore explore the possibility of replacing the missing values 
(data imputation, e.g. \citealt{lpw21,gnd+22}), 
via the best two performing methods  found by \citet{cur22}:
\begin{enumerate} 
\item {\it Simple imputation:} The missing feature is replaced with a constant, either the mean, 
median or most frequently occuring value. All of these methods were effective for training and validation upon the SDSS sample,
in particular replacing the missing magnitudes with the maximum value for that band  (analogous to assuming that the missing
magnitude is at the detection limit, \citealt{cma+21}). This method, however, was not effective in applying
the SDSS model to the imputed radio samples.
\item {\it Multivariate  (multiple)  imputation:} Machine learning is used to estimate the
missing values from the other features.\footnote{Via  the {\tt IterativeImputer} function 
of {\sf sklearn} (https://scikit-learn.org/stable/), where, again, we 
remove the spectroscopic redshift prior to the imputation so that it is not used as a feature  (see \citealt{cur22} for details).}
Although, along with the other model-based methods which were tested, this is not as good a performer as simple imputation 
when validating upon the same dataset,  it did produce the best results when used to impute the missing values in the three radio samples,
which were trained by the SDSS model.
If the number of missing magnitudes was limited to two per source high quality predictions were retained  \citep{cur22}.
\end{enumerate} 

\begin{figure*}
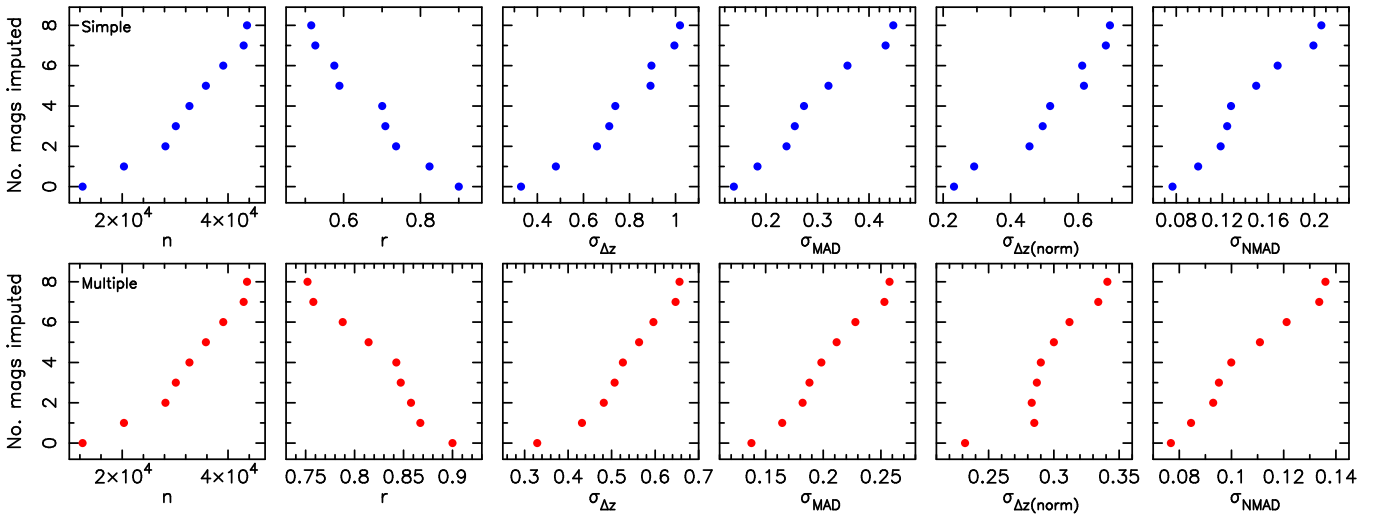

\centering \includegraphics[angle=-90,scale=0.61]{6-max_SDSS-single.eps}
\centering \includegraphics[angle=-90,scale=0.61]{6-multi_SDSS-single.eps}
\caption{The performance of the ANN for different numbers of imputed magnitudes per each of the Milliquas sources,
using simple (top) and multivariate (bottom) imputation, predicted from the SDSS model. 
The number of sources ranges from 12\,503 (no imputed magnitudes) to 43\,595 (99\% of the sample for eight imputed).}
\label{6-SDSS}
\end{figure*}

As previously, for training and validating on the same dataset, we find replacing the missing magnitudes 
with the maximum value for that band
\begin{figure*}
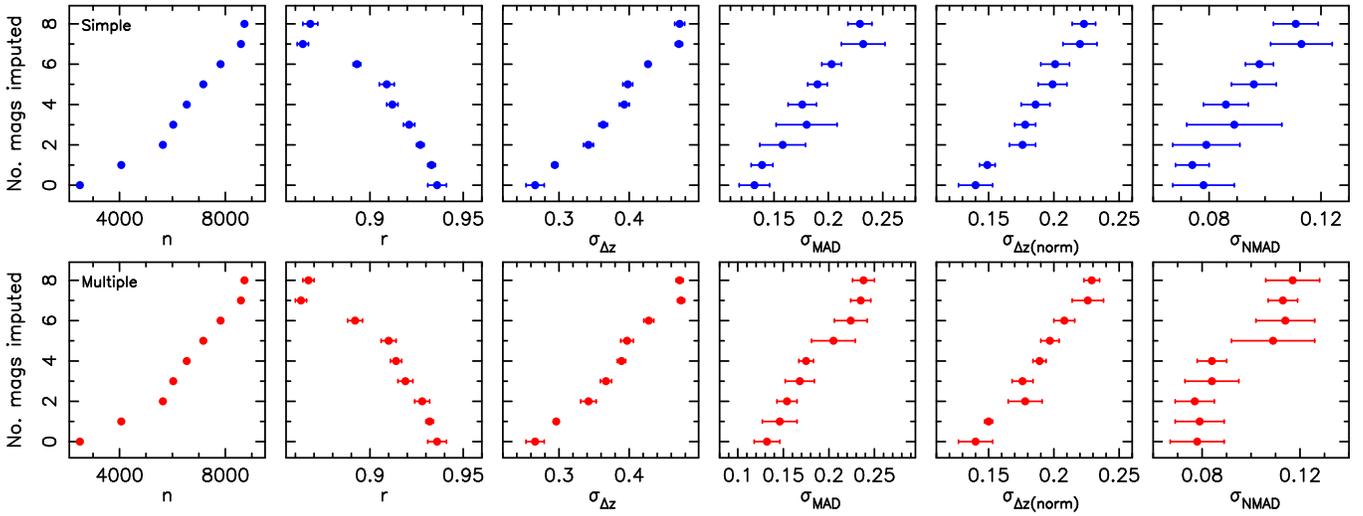

\centering \includegraphics[angle=-90,scale=0.61]{6-max_self-row.eps}
\centering \includegraphics[angle=-90,scale=0.61]{6-multi_self-row.eps}
\caption{The performance of the ANN for different numbers of imputed magnitudes per source
using simple  (top) and multivariate (bottom) imputation for self-training with the Milliquas sources. 
The 20\% self-validation number ranges from 2501 (no imputed magnitudes) to 8719 (eight imputed).
The error bars show the $\pm1\sigma$ from the mean of 10 trials, each with a randomly selected training 
and validation sub-sample.}
\label{6-self}
\end{figure*} 
to be  an effective imputation strategy (Fig.~\ref{6-self}, top), with the results indistinguishable from the multivariate
imputation (Fig.~\ref{6-self}, bottom). However, due to the majority of the sample being used for training, this
only yields predictions for a small portion of the sample. Therefore in Fig.~\ref{6-SDSS} we show the results of
validating upon the imputed Milliquas data trained on the SDSS data (see Sect.~\ref{stms}).
From this, we see that, unlike when self-training, the multiple imputation significantly outperforms the simple imputation.
From the metrics it is not clear what a reasonable limit to the number of  imputed magnitudes should be, although 
the regression coefficient (and perhaps NMAD) suggests up to four, while the normalised standard deviation suggests that this 
may be best limited to one.

\begin{figure}
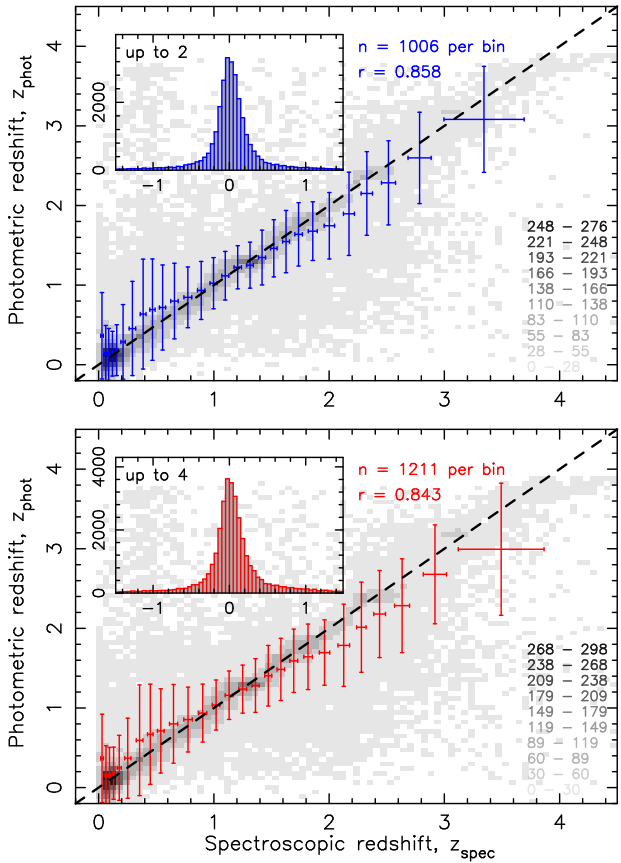
 
\centering \includegraphics[angle=-90,scale=0.5]{pred_X-SDSS_imp2_1006.eps}
\centering \includegraphics[angle=-90,scale=0.5]{pred_X-SDSS_imp4_1211.eps}
\caption{The photometric versus the spectroscopic redshift for the Milliquas radio sources from the SDSS model
(Fig.~\ref{pred_X-SDSS}), but where up to two (top) and four (bottom) missing values per source are imputed. 
In the top panel, $n=28\,173$, $\sigma_{\Delta z} = 0.482$,  $\sigma_{\text{MAD}}=0.182$, $\sigma_{\Delta z(\text{norm})} = 0.283$, $\sigma_{\text{NMAD}}=0.093$ and
$\eta=24.5$\%. In the bottom panel, $n=32\,698$, $\sigma_{\Delta z} = 0.526$,  $\sigma_{\text{MAD}}=0.198$, $\sigma_{\Delta z(\text{norm})} = 0.290$ $\sigma_{\text{NMAD}}=1.00$ and $\eta=26.8$\%.}.
\label{pred_X-imp}
\end{figure}

Replacing the missing data for sources with only one absent magnitude, gives only a moderate increase in 
sample size (to 20\,332) and \citet{cur22} suggests that up to two is reasonable when applying the SDSS model
to external datasets. Therefore in Fig.~\ref{pred_X-imp}, we show the results for both up to two and four imputed
magnitudes per source.
Compared to the  12\,503 non-imputed sources (Fig.~\ref{pred_X-SDSS}), the binned data clearly shows
in increasing inaccuracy in the photometric redshift predictions, with the 
outlier fraction climbing from $\eta=19.0$\%  for 12\,504 sources to $24.5$\%  for 28\,173 and  $26.8$\% for 32\,698.
Thus, the number of sources which can be fit by imputing up to four features per sources is 2.6 times that
of the sources with all of the magnitudes (up from 28\% to 74\% of the sample), at the expense of 1.4 times as
many outliers.

\begin{figure} 
\centering \includegraphics[angle=-90,scale=0.5]{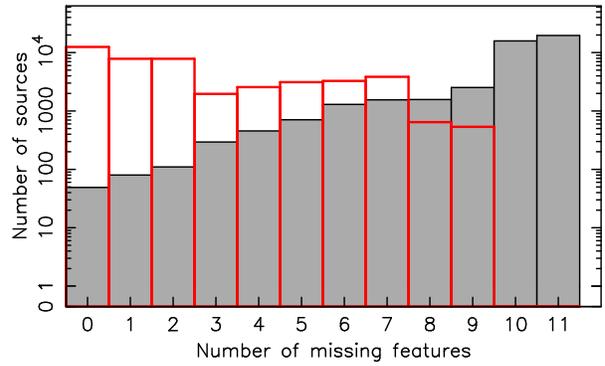}
\caption{The number of missing radio flux measurements (filled histogram) and the number of missing
magnitudes (unfilled) for the 41\,119 Milliquas quasars.} 
\label{missing_feats}
\end{figure}

Given the effectiveness of the imputation of the missing magnitudes, we also explored the possibility of imputing
the missing radio data, adding much-needed training data  (cf. Table~\ref{intersect_all}).  In order
to minimise the number of sources with many missing features and utilise numerical values only, we
considered only the  {\sc flux} sample. Even so, unlike the NIR--optical--UV case, the vast majority of sources 
still have a large number of missing features (Fig.~\ref{missing_feats}).
For example, imputing up to two missing values per source only yields 239 sources, with the imputation
of six required to reach 2990. This nearly doubles (to 5957) if we exclude the 15 and 20~GHz flux densities, 
although, perhaps unsurprisingly, there is no improvement in the machine learning results ($r\approx0.29, \,\sigma_{\Delta z} \approx0.81,\, \sigma_{\text{MAD}}\approx0.82,\, \sigma_{\Delta z(\text{norm})} \approx 0.42,\,\sigma_{\text{NMAD}}\approx0.37$ 
and  $\eta\approx70$\%). We therefore conclude that,  due to the majority of sources being dominated
by a large number of missing features, data imputation is not (yet) useful for replacing the missing radio data.

\section{Discussion}

\subsection{Radio photometric redshifts}

Despite extensive testing, for which the main models are described in Sect.~\ref{rpr}, we 
could not find a machine learning model or  artificial neutral network
which could accurately predict the redshifts of the
sources based upon features in the radio SED alone. Although the training sample is significantly
smaller, $\approx1000$ sources  compared to $\approx10\,000$ for the Milliquas validation sample,
the NIR--optical--UV photometry can still produce accurate predictions for the smaller sample size.
It should be noted, however, that these 1000 sources have the full NIR--optical--UV photometry, comprising
nine features each, while the radio sample has just six features (including the presence of a turnover 
in the SED, Table~\ref{intersect_all}). We also find circumstantial evidence that a much larger radio photometric sample may 
yet prove useful (Fig.~\ref{frac_test}).

Using the radio flux densities at different frequencies is analogous  to the magnitudes 
in the prediction of redshifts, although  from Fig.~\ref{radio_plots} it is apparent that the radio
photometry may be steeply  flux limited, a clear example being the abrupt
decrease in sources with $S_{1.4~\text{GHz}}\lapp10^{-3}$~Jy. 
Over a heterogeneously sampled
radio SED this could cause some frequencies to drop out of the training model before others, causing
uneven sampling of the sources. 
This may also be responsible for the bimodal distribution in the flux densities, which are not correlated
strongly with redshift (e.g.  $S_{\text{150 MHz}}$, Table~\ref{z_stats}).
This of course could also be an issue for the NIR--optical--UV magnitudes,
but each of the these bands are dominated by a single survey (WISE, SDSS \& GALEX) and the SED as a 
whole is dominated by the  SDSS.

\begin{figure}
\centering \includegraphics[angle=-90,scale=0.5]{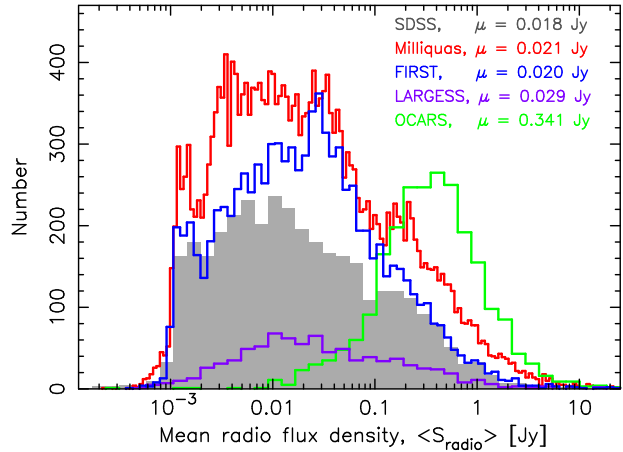}
\caption{The mean radio flux densities of the QSOs/quasars in the SDSS, Milliquas, FIRST, 
LARGESS and OCARS with measured $\leq10$~GHz flux densities (updated from \citealt{cmp21}). 
In order to fit within a similar range,  the Milliquas binning is four times finer and the 
FIRST twice as fine as for the other samples.}
\label{radio_hist}
\end{figure}

In Fig.~\ref{radio_hist}, we see that, like the SDSS and FIRST samples, 
Milliquas does appear to suffer from a flux limitation.
This is least  apparent for the  OCARS fluxes, probably due to the much higher brightness of these
sources.  The lack of apparent flux limitation suggests that OCARS may provide a better model
for the radio photometric redshift. Testing this, we find the results to also be disappointing,
although the sample significantly smaller than the Milliquas  (see Appendix A).

We note that direct comparison of the radio properties with the NIR--optical--UV magnitudes
may not be justified, as the latter spans over three observing bands, although arguably the 70~MHz -- 20~GHz
is also over several bands, even if unevenly sampled. Other previous studies have also failed to 
find a photometric redshift \citep{maj15,nsl+19}, a possible explanation being the relatively featureless
radio SEDs: While the value of turnover frequency may be dictated by intrinsic source properties such as the extent 
of the radio emission (e.g. \citealt{ode98,fan00,omd06}) and the electron density 
\citep{dbo97,chj+19}, the observed wavelengths of the  $\lambda\sim1~\mu$m inflection in the 
NIR and $\lambda= 1216$~\AA\ Lyman-break in the UV are two
features which are dependent on the  redshift.

\subsection{NIR--optical--UV photometric redshifts}

Although a redshift prediction from radio data is not yet possible, we find that
the redshifts of sources in the external  three radio catalogues can be accurately predicted 
from a model trained on the Milliquas radio sources  using their NIR--optical--UV magnitudes.
This was previously shown to be the case for training on a sample of SDSS QSOs \citep{cmp21} 
and, the fact that the Milliquas give very similar metrics  (Sect.~\ref{stms}), gives us confidence that {\em SkyMapper} 
will be able to provide a model for the southern sources, not accessible to the SDSS (see Sect.~\ref{imp}).

\begin{figure}
\centering \includegraphics[angle=-90,scale=0.5]{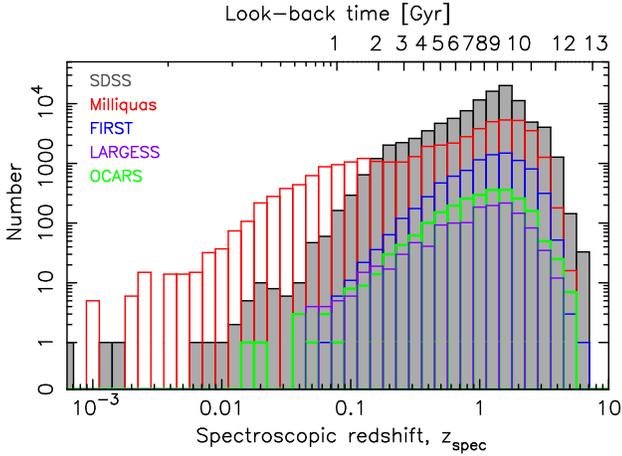}
\caption{The redshift distributions of the different samples.}
\label{z_histo}
\end{figure}

\begin{figure}
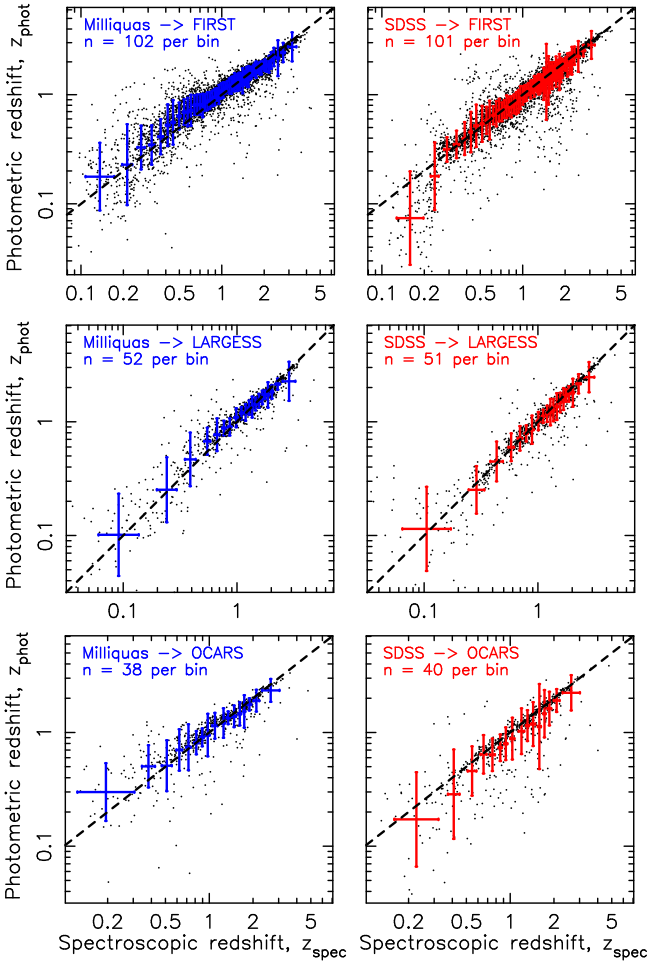
 
\centering \includegraphics[angle=-90,scale=0.5]{log_X-FIRST_102.eps}
\centering \includegraphics[angle=-90,scale=0.5]{log_X-LARGESS_52.eps}
\centering \includegraphics[angle=-90,scale=0.5]{log_X-OCARS_38.eps}
\caption{Comparison of the Milliquas (left) and SDSS (right) models emphasising the low redshifts. Top: Redshift predictions for the 
FIRST QSOs versus the measured redshifts. Middle: LARGESS QSOs. Bottom: OCARS quasars. The bin sizes differ slightly as the
very few $z_{\text{phot}}<0$ predictions are excluded.}
\label{log_X}
\end{figure}

This similarity in the results of the Milliquas and SDSS models arises despite the very different magnitude 
(Fig.~\ref{SDSS_histos}) and redshift (Fig.~\ref{z_histo})  distributions.
From the latter, we see that the Milliquas radio sources sample the low redshifts ($z\lapp0.1$) significantly more
comprehensively than our SDSS training sample. This may result in the Milliquas  providing a better
model at low redshift. 
Examining this, in Fig.~\ref{log_X}
we see that any improvement at low redshift is most pronounced for the FIRST predictions at $z\lapp0.2$.
We also see that overall the Milliquas does appear
to provide a better model of the OCARS redshift predictions. From the statistics (Table~\ref{SDSS_Miiliqua}), 
the Milliquas does rate significantly better according to the regression coefficient and 
standard deviations, although the median absolute deviations are poorer to a similar degree.\footnote{For $\sigma_{\Delta z (\text{norm})}$, $(0.170-0.131)/1.70 = 0.23$, cf.  $\sigma_{\text{NMAD}}$, $(0.063-0.076)/0.063 = -0.21$.}

\section{Conclusions}

In order to expand upon our previous work in using machine learning and artificial neural networks   to obtain the 
redshifts of radio sources, we have explored using the {\em Million Quasars Catalogue}. From this,
there are 44\,495 sources classed as radio sources which have  a spectroscopic redshift. We were
able to match 44\,119 of these in NED and obtain the available photometry, spanning the radio to
ultraviolet bands. The aim was to obtain a large sample of quasars with which to test for potential 
redshift prediction from the radio spectrum alone. For these we test:
\begin{itemize}
  \item  Peaked spectrum sources only, where the turnover parameters are included as features.
\item  All of the sources, by one-hot encoding the presence of a turnover in the spectrum as a feature.
 \item  All of the sources, using only the flux densities as features.
\end{itemize}
We find from these models that the radio spectrum has little predictive power. One possibility is the trade-off between
sample size and number of features, although testing the NIR--optical--UV model on a similar size sample shows a sample
of $\sim1000$ to still be useful. However, it may still be possible for a much larger sample to yield positive results
from the largely featureless radio SEDs, especially if the rare features (e.g. the 70 \& 150~MHz fluxes) prove
to be critical, and we do find marginal evidence for the predictive power of the models improving with training 
sample size.  

Another cause of the poor performance may be the flux limitation, where the sample is abruptly truncated at flux
densities of $S\lapp10^{-3}$~Jy.  For fluxes compiled from several disparate sources, this would introduce non-uniform
sampling of the SEDs, effectively scrambling the data. This flux limitation is not apparent for one of our test samples,
the OCARS catalogue of VLBI astrometry sources, probably due to the much higher flux densities. Again, however, this has
little predictive power although the amount of available training data is significantly smaller.

We also explore how the NIR--optical--UV photometry of the 12\,503  Milliquas radio sources, with the full 
$W2, W1, z,i,r,g,u,NUV,FUV$ complement, compares to that of the SDSS in providing a model with
which to predict of the redshifts of radio selected samples:
\begin{enumerate}
  \item Despite differences in their distributions, the same NIR--optical--UV photometry, which has
    a proven success in yielding photometric redshifts for  SDSS QSOs, appears to work equally well in the training and 
    validation of the Milliquas radio sources. Training on 10\,002 and validating on 2501 of the
    sample, gives an outlier fraction of $\eta = 12\%$. 
  
    \item Training a model on the SDSS sample \citep{cmp21} and validating on the 12\,503 
      Milliquas sources is successful, with  an outlier fraction of  $\eta = 19\%$.  

      \item Training a model on the Milliquas sample and validating on three external 
        catalogues of radio selected sources yields outlier fractions of $\eta\leq18$\% and
        spreads in $\Delta z$ which are indistinguishable from training on the SDSS data.
        This transferability gives us confidence that  future {\em SkyMapper} measurements of the 
        $u, v, g, r, i, z$  bands in the southern sky can be combined with the NIR and 
        UV photometry to give a reliable model to yield photometric redshifts of sources detected with the 
        SKA.

      \item  In some cases Milliquas outperforms the SDSS model in the validation of the 
        external radio sources at low redshifts due
        to better sampling. This suggests that the results from different training models may be reliably combined
        according to where they have the best redshift coverage.

\item As per the SDSS data \citep{cur22}, machine learning can be used to impute the missing data
  with the replacement of up to four magnitudes per source, increasing the number of sources to 32\,698
(73\% of the sample) at the cost of the outlier fraction rising to 27\%.

\end{enumerate}
To conclude, although a radio photometric redshift remains elusive, the fact that the same optical
bands of a distinct dataset can provide an accurate model 
promises great potential in using machine learning to obtain the redshifts of the large number
of sources to be discovered in forthcoming surveys.

\section*{Acknowledgements}

We wish to thank the referee for their  helpful and detailed feedback which helped improve the manuscript.
This research has made use of the NASA/IPAC Extragalactic
Database (NED) which is operated by the Jet Propulsion Laboratory, California Institute of Technology, under contract
with the National Aeronautics and Space Administration and NASA's Astrophysics Data System Bibliographic Service. This
research has also made use of NASA's Astrophysics Data System Bibliographic Service.
Funding for the SDSS has been provided by the Alfred P. Sloan Foundation, the Participating Institutions, the National Science Foundation, the U.S. Department of Energy, the National Aeronautics and Space Administration, the Japanese Monbukagakusho, the Max Planck Society, and the Higher Education Funding Council for England. 
 This publication makes use of data products from the Wide-field Infrared Survey
Explorer, which is a joint project of the University of California, Los Angeles, and the Jet Propulsion
Laboratory/California Institute of Technology, funded by the National Aeronautics and Space Administration.  This
publication makes use of data products from the Two Micron All Sky Survey, which is a joint project of the University of
Massachusetts and the Infrared Processing and Analysis Center/California Institute of Technology, funded by the National
Aeronautics and Space Administration and the National Science Foundation. GALEX is operated for NASA by the California
Institute of Technology under NASA contract NAS5-98034.  

\section*{Data availability} 

Data and training models available on request.


\begin{thebibliography}{46}
\expandafter\ifx\csname natexlab\endcsname\relax\def\natexlab#1{#1}\fi

\bibitem[{{Athreya} \& {Kapahi}(1998)}]{ak98}
{Athreya} R.~M., {Kapahi} V.~K., 1998, JA\&A, 19, 63

\bibitem[{Becker {et~al}\mbox{.}(1995)Becker, White, \& Helfand}]{bwh95}
Becker R.~H., White R.~L., Helfand D.~J., 1995, ApJ, 450, 559

\bibitem[{{Bianchi} {et~al}\mbox{.}(2014){Bianchi}, {Conti}, \&
  {Shiao}}]{bcs14}
{Bianchi} L., {Conti} A., {Shiao} B., 2014, Advances in Space Research, 53, 900

\bibitem[{{Bianchi} {et~al}\mbox{.}(2017){Bianchi}, {Shiao}, \&
  {Thilker}}]{bst17}
{Bianchi} L., {Shiao} B., {Thilker} D., 2017, ApJS, 230, 24

\bibitem[{{Bovy} {et~al}\mbox{.}(2012){Bovy}, {Myers}, {Hennawi}, {Hogg},
  {McMahon}, {Schiminovich}, {Sheldon}, {Brinkmann}, {Schneider}, \&
  {Weaver}}]{bmh+12}
{Bovy} J. {et~al.}, 2012, ApJ, 749, 41

\bibitem[{{Brescia} {et~al}\mbox{.}(2013){Brescia}, {Cavuoti}, {D'Abrusco},
  {Longo}, \& {Mercurio}}]{bcd+13}
{Brescia} M., {Cavuoti} S., {D'Abrusco} R., {Longo} G., {Mercurio} A., 2013,
  ApJ, 772, 140

\bibitem[{{Bridle} {et~al}\mbox{.}(1972){Bridle}, {Kesteven}, \&
  {Guindon}}]{bkg72}
{Bridle} A.~H., {Kesteven} M.~J.~L., {Guindon} B., 1972, Astrophysical Letters,
  11, 27

\bibitem[{{Carvajal} {et~al}\mbox{.}(2021){Carvajal}, {Matute}, {Afonso},
  {Amarantidis}, {Barbosa}, {Cunha}, \& {Humphrey}}]{cma+21}
{Carvajal} R., {Matute} I., {Afonso} J., {Amarantidis} S., {Barbosa} D.,
  {Cunha} P., {Humphrey} A., 2021, A New Window on the Radio Emission from
  Galaxies, Galaxy Clusters and Cosmic Web: Current Status and Perspectives

\bibitem[{{Ching} {et~al}\mbox{.}(2017){Ching}, {Sadler}, {Croom}, {Johnston},
  {Pracy}, {Couch}, {Hopkins}, {Jurek}, \& {Pimbblet}}]{csc+17}
{Ching} J.~H.~Y. {et~al.}, 2017, MNRAS, 464, 1306

\bibitem[{Curran {et~al}\mbox{.}(2013)Curran, {Whiting}, {Sadler}, \&
  {Bignell}}]{cwsb12}
Curran S., {Whiting} M.~T., {Sadler} E.~M., {Bignell} C., 2013, MNRAS, 428,
  2053

\bibitem[{Curran(2020)}]{cur20}
Curran S.~J., 2020, MNRAS, 493, L70

\bibitem[{Curran(2021)}]{cur21b}
Curran S.~J., 2021, MNRAS, 508, 1165

\bibitem[{Curran(2022)}]{cur22}
Curran S.~J., 2022, MNRAS, 512, 2099

\bibitem[{Curran {et~al}\mbox{.}(2019)Curran, {Hunstead}, {Johnston},
  {Whiting}, {Sadler}, {Allison}, \& {Athreya}}]{chj+19}
Curran S.~J., {Hunstead} R.~W., {Johnston} H.~M., {Whiting} M.~T., {Sadler}
  E.~M., {Allison} J.~R., {Athreya} R., 2019, MNRAS, 484, 1182

\bibitem[{Curran \& Moss(2019)}]{cm19}
Curran S.~J., Moss J.~P., 2019, A\&A, 629, A56

\bibitem[{Curran {et~al}\mbox{.}(2021)Curran, Moss, \& Perrott}]{cmp21}
Curran S.~J., Moss J.~P., Perrott Y.~C., 2021, MNRAS, 503, 2639

\bibitem[{Curran \& Whiting(2012)}]{cw12}
Curran S.~J., Whiting M.~T., 2012, ApJ, 759, 117

\bibitem[{Curran {et~al}\mbox{.}(2011)Curran, Whiting, Combes, Kuno, Francis,
  Nakai, Webb, Murphy, \& Wiklind}]{cwc+11}
Curran S.~J. {et~al.}, 2011, MNRAS, 416, 2143

\bibitem[{{de Vries} {et~al}\mbox{.}(1997){de Vries}, {Barthel}, \&
  {O'Dea}}]{dbo97}
{de Vries} W.~H., {Barthel} P.~D., {O'Dea} C.~P., 1997, A\&A, 321, 105

\bibitem[{{Fanti}(2000)}]{fan00}
{Fanti} C., 2000, in EVN Symposium 2000, Proceedings of the 5th European VLBI
  Network Symposium, {Conway} J.~E., {Polatidis} A.~G., {Booth} R.~S.,
  {Pihlstr{\"o}m} Y.~M., eds., Onsala Space Observatory, Chalmers Technical
  University, G\"{o}teborg, Sweden, p.~73

\bibitem[{{Flesch}(2015)}]{fle15}
{Flesch} E.~W., 2015, PASA, 32, 1

\bibitem[{{Flesch}(2021)}]{fle21}
{Flesch} E.~W., 2021, arXiv e-prints, arXiv:2105.12985

\bibitem[{{Gibson} {et~al}\mbox{.}(2022){Gibson}, {Narendra}, {Dainotti},
  {Bogdan}, {Pollo}, {Poliszczuk}, {Rinaldi}, \& {Liodakis}}]{gnd+22}
{Gibson} S.~J., {Narendra} A., {Dainotti} M.~G., {Bogdan} M., {Pollo} A.,
  {Poliszczuk} A., {Rinaldi} E., {Liodakis} I., 2022, Frontiers in Astronomy
  and Space Sciences, arXiv:2203.00087

\bibitem[{{Han} {et~al}\mbox{.}(2016){Han}, {Ding}, {Zhang}, \&
  {Zhao}}]{hdzz16}
{Han} B., {Ding} H.-P., {Zhang} Y.-X., {Zhao} Y.-H., 2016, Research in
  Astronomy and Astrophysics, 16, 74

\bibitem[{{Li} {et~al}\mbox{.}(2021){Li}, {Zhang}, {Cui}, {Fan}, {Zhao}, {Wu},
  {Zhang}, {Han}, {Xu}, {Tao}, {Li}, \& {He}}]{lzc+21}
{Li} C. {et~al.}, 2021, MNRAS, 509, 2289

\bibitem[{{Luken} {et~al}\mbox{.}(2021){Luken}, {Padhy}, \& {Wang}}]{lpw21}
{Luken} K.~J., {Padhy} R., {Wang} X.~R., 2021, in Machine Learning for Physical
  Sciences workshop at NeurIPS 2021

\bibitem[{{Ma} {et~al}\mbox{.}(2009){Ma}, {Arias}, {Bianco}, {Boboltz},
  {Bolotin}, {Charlot}, {Engelhardt}, {Fey}, {Gaume}, {Gontier}, {Heinkelmann},
  {Jacobs}, {Kurdubov}, {Lambert}, {Malkin}, {Nothnagel}, {Petrov},
  {Skurikhina}, {Sokolova}, {Souchay}, {Sovers}, {Tesmer}, {Titov}, {Wang},
  {Zharov}, {Barache}, {Boeckmann}, {Collioud}, {Gipson}, {Gordon}, {Lytvyn},
  {MacMillan}, \& {Ojha}}]{mab+09}
{Ma} C. {et~al.}, 2009, IERS Technical Note, 35, 1

\bibitem[{{Maddox} {et~al}\mbox{.}(2012){Maddox}, {Hewett}, {P{\'e}roux},
  {Nestor}, \& {Wisotzki}}]{mhp+12}
{Maddox} N., {Hewett} P.~C., {P{\'e}roux} C., {Nestor} D.~B., {Wisotzki} L.,
  2012, MNRAS, 424, 2876

\bibitem[{Majic \& Curran(2015)}]{maj15}
Majic R. A.~M., Curran S.~J., 2015, {Radio Photometric Redshifts: Estimating
  radio source redshifts from their spectral energy distributions}. Tech. rep.,
  Victoria University of Wellington

\bibitem[{{Malkin}(2018)}]{mal18}
{Malkin} Z., 2018, ApJS, 239, 20

\bibitem[{{Menon}(1983)}]{men83}
{Menon} T.~K., 1983, AJ, 88, 598

\bibitem[{Morganti {et~al}\mbox{.}(2015)Morganti, Sadler, \& Curran}]{msc+15}
Morganti R., Sadler E.~M., Curran S., 2015, Advancing Astrophysics with the
  Square Kilometre Array (AASKA14), 134

\bibitem[{{Nakoneczny} {et~al}\mbox{.}(2021){Nakoneczny}, {Bilicki}, {Pollo},
  {Asgari}, {Dvornik}, {Erben}, {Giblin}, {Heymans}, {Hildebrandt},
  {Kannawadi}, {Kuijken}, {Napolitano}, \& {Valentijn}}]{nbp+21}
{Nakoneczny} S.~J. {et~al.}, 2021, A\&A, 649, A81

\bibitem[{{Norris} {et~al}\mbox{.}(2011){Norris}, {Hopkins}, {Afonso}, {Brown},
  {Condon}, {Dunne}, {Feain}, {Hollow}, {Jarvis}, {Johnston-Hollitt}, {Lenc},
  {Middelberg}, {Padovani}, {Prandoni}, {Rudnick}, {Seymour}, {Umana},
  {Andernach}, {Alexander}, {Appleton}, {Bacon}, {Banfield}, {Becker}, {Brown},
  {Ciliegi}, {Jackson}, {Eales}, {Edge}, {Gaensler}, {Giovannini}, {Hales},
  {Hancock}, {Huynh}, {Ibar}, {Ivison}, {Kennicutt}, {Kimball}, {Koekemoer},
  {Koribalski}, {L{\'o}pez-S{\'a}nchez}, {Mao}, {Murphy}, {Messias},
  {Pimbblet}, {Raccanelli}, {Randall}, {Reiprich}, {Roseboom},
  {R{\"o}ttgering}, {Saikia}, {Sharp}, {Slee}, {Smail}, {Thompson}, {Urquhart},
  {Wall}, \& {Zhao}}]{nha+11}
{Norris} R.~P. {et~al.}, 2011, PASA, 28, 215

\bibitem[{{Norris} {et~al}\mbox{.}(2019){Norris}, {Salvato}, {Longo},
  {Brescia}, {Budavari}, {Carliles}, {Cavuoti}, {Farrah}, {Geach}, {Luken},
  {Musaeva}, {Polsterer}, {Riccio}, {Seymour}, {Smol{\v c}i{\'c}}, {Vaccari},
  \& {Zinn}}]{nsl+19}
{Norris} R.~P. {et~al.}, 2019, PASP, 131, 108004

\bibitem[{{O'Dea}(1998)}]{ode98}
{O'Dea} C.~P., 1998, PASP, 110, 493

\bibitem[{{O'Dea} \& {Baum}(1997)}]{ob97}
{O'Dea} C.~P., {Baum} S.~A., 1997, AJ, 113, 148

\bibitem[{{Orienti} {et~al}\mbox{.}(2006){Orienti}, {Morganti}, \&
  {Dallacasa}}]{omd06}
{Orienti} M., {Morganti} R., {Dallacasa} D., 2006, A\&A, 457, 531

\bibitem[{{P{\^a}ris} {et~al}\mbox{.}(2018){P{\^a}ris}, {Petitjean}, {Aubourg},
  {Myers}, {Streblyanska}, {Lyke}, {Anderson}, {Armengaud}, {Bautista},
  {Blanton}, {Blomqvist}, {Brinkmann}, {Brownstein}, {Brandt}, {Burtin},
  {Dawson}, {de la Torre}, {Georgakakis}, {Gil-Mar{\'\i}n}, {Green}, {Hall},
  {Kneib}, {LaMassa}, {Le Goff}, {MacLeod}, {Mariappan}, {McGreer}, {Merloni},
  {Noterdaeme}, {Palanque-Delabrouille}, {Percival}, {Ross}, {Rossi},
  {Schneider}, {Seo}, {Tojeiro}, {Weaver}, {Weijmans}, {Y{\`e}che}, {Zarrouk},
  \& {Zhao}}]{ppa+18}
{P{\^a}ris} I. {et~al.}, 2018, A\&A, 613, A51

\bibitem[{{Richards} {et~al}\mbox{.}(2001){Richards}, {Weinstein}, {Schneider},
  {Fan}, {Strauss}, {Vanden Berk}, {Annis}, {Burles}, {Laubacher}, {York},
  {Frieman}, {Johnston}, {Scranton}, {Gunn}, {Ivezi{\'c}}, {Nichol},
  {Budav{\'a}ri}, {Csabai}, {Szalay}, {Connolly}, {Szokoly}, {Bahcall},
  {Ben{\'{\i}}tez}, {Brinkmann}, {Brunner}, {Fukugita}, {Hall}, {Hennessy},
  {Knapp}, {Kunszt}, {Lamb}, {Munn}, {Newberg}, \& {Stoughton}}]{rws+01}
{Richards} G.~T. {et~al.}, 2001, AJ, 122, 1151

\bibitem[{{Skrutskie} {et~al}\mbox{.}(2006){Skrutskie}, {Cutri}, {Stiening},
  {Weinberg}, {Schneider}, {Carpenter}, {Beichman}, {Capps}, {Chester},
  {Elias}, {Huchra}, {Liebert}, {Lonsdale}, {Monet}, {Price}, {Seitzer},
  {Jarrett}, {Kirkpatrick}, {Gizis}, {Howard}, {Evans}, {Fowler}, {Fullmer},
  {Hurt}, {Light}, {Kopan}, {Marsh}, {McCallon}, {Tam}, {Van Dyk}, \&
  {Wheelock}}]{scs+06}
{Skrutskie} M.~F. {et~al.}, 2006, AJ, 131, 1163

\bibitem[{{Snellen} {et~al}\mbox{.}(1998){Snellen}, {Schilizzi}, {de Bruyn},
  {Miley}, {Rengelink}, {Roettgering}, \& {Bremer}}]{ssd+98}
{Snellen} I.~A.~G., {Schilizzi} R.~T., {de Bruyn} A.~G., {Miley} G.~K.,
  {Rengelink} R.~B., {Roettgering} H.~J., {Bremer} M.~N., 1998, A\&AS, 131, 435

\bibitem[{{Turner} {et~al}\mbox{.}(2020){Turner}, {Drouart}, {Seymour}, \&
  {Shabala}}]{tdss20}
{Turner} R.~J., {Drouart} G., {Seymour} N., {Shabala} S.~S., 2020, MNRAS, 499,
  3660

\bibitem[{{Weinstein} {et~al}\mbox{.}(2004){Weinstein}, {Richards},
  {Schneider}, {Younger}, {Strauss}, {Hall}, {Budav{\'a}ri}, {Gunn}, {York}, \&
  {Brinkmann}}]{wrs+04}
{Weinstein} M.~A. {et~al.}, 2004, ApJS, 155, 243

\bibitem[{{White} {et~al}\mbox{.}(1997){White}, {Becker}, {Helfand}, \&
  {Gregg}}]{wbhg97}
{White} R.~L., {Becker} R.~H., {Helfand} D.~J., {Gregg} M.~D., 1997, ApJ, 475,
  479

\bibitem[{{Wolf} {et~al}\mbox{.}(2018){Wolf}, {Onken}, {Luvaul}, {Schmidt},
  {Bessell}, {Chang}, {Da Costa}, {Mackey}, {Martin-Jones}, {Murphy},
  {Preston}, {Scalzo}, {Shao}, {Smillie}, {Tisserand}, {White}, \&
  {Yuan}}]{wol+18}
{Wolf} C. {et~al.}, 2018, PASA, 35, 10

\end{thebibliography}

\label{lastpage}

\section*{Appendix A}
\label{appa}

\subsection*{Radio photometric redshifts from OCARS}

\begin{figure*}
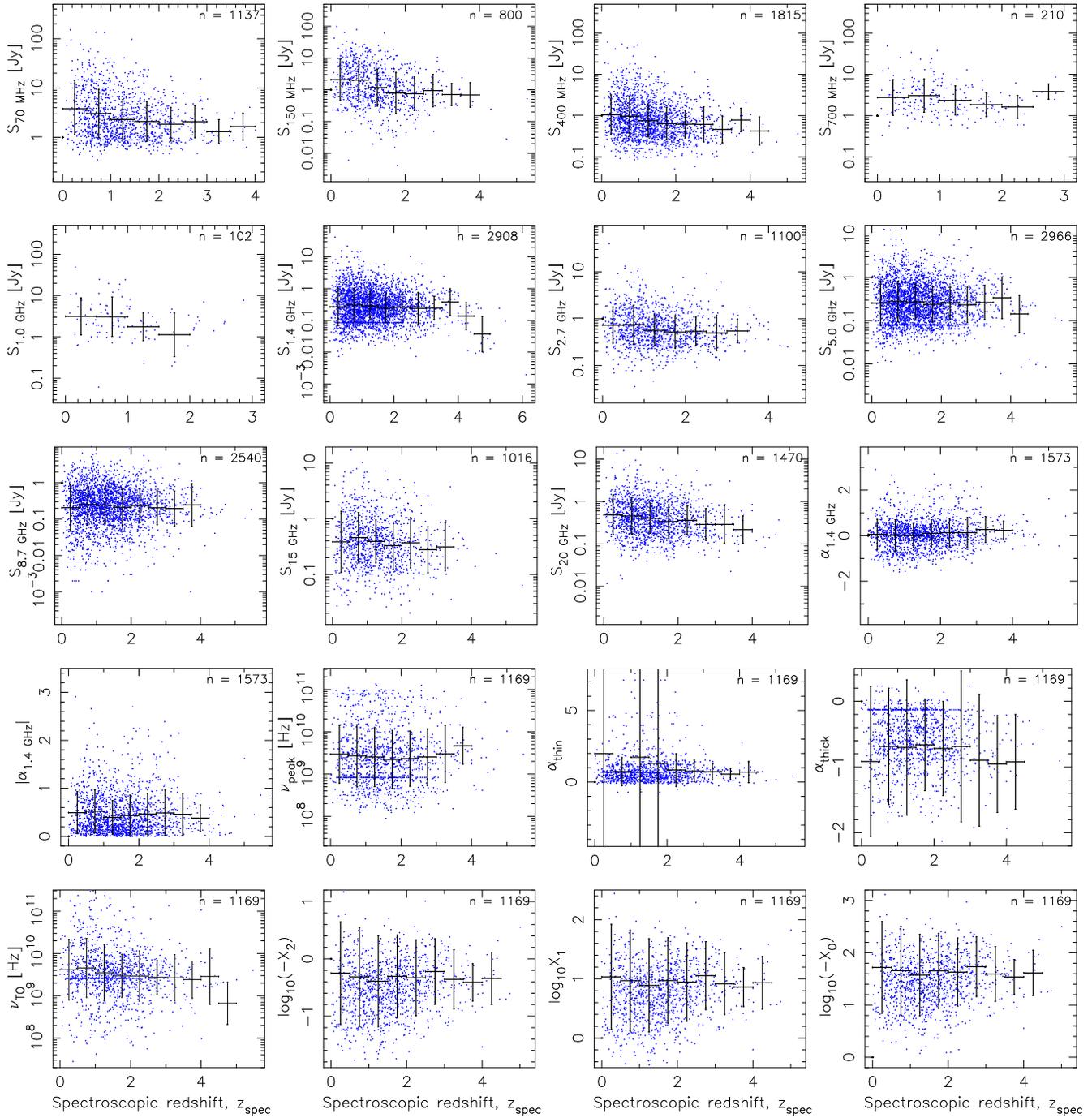

\centering \includegraphics[angle=-90,scale=0.26]{S_70-OCARS_113.eps}
\centering \includegraphics[angle=-90,scale=0.26]{S_150-OCARS_80.eps}
\centering \includegraphics[angle=-90,scale=0.26]{S_400-OCARS_181.eps}
\centering \includegraphics[angle=-90,scale=0.26]{S_700-OCARS_21.eps}
\centering \includegraphics[angle=-90,scale=0.26]{S_1-OCARS_25.eps}
\centering \includegraphics[angle=-90,scale=0.26]{S_1p4-OCARS_290.eps}
\centering \includegraphics[angle=-90,scale=0.26]{S_2p7_OCARS_110.eps}
\centering \includegraphics[angle=-90,scale=0.26]{S_5-OCARS_296.eps}
\centering \includegraphics[angle=-90,scale=0.26]{S_8p7-OCARS_254.eps}
\centering \includegraphics[angle=-90,scale=0.26]{S_15-OCARS_127.eps}
\centering \includegraphics[angle=-90,scale=0.26]{S_20-OCARS_147.eps}
\centering \includegraphics[angle=-90,scale=0.26]{SI_OCARS_157.eps}
\centering \includegraphics[angle=-90,scale=0.26]{SI-abs_OCARS_157.eps}
\centering \includegraphics[angle=-90,scale=0.26]{TO_ssd+98-OCARS_116.eps}
\centering \includegraphics[angle=-90,scale=0.26]{thin_OCARS_116.eps}
\centering \includegraphics[angle=-90,scale=0.26]{thick_OCARS_116.eps}
\centering \includegraphics[angle=-90,scale=0.26]{TO_poly-OCARS_116.eps}
\centering \includegraphics[angle=-90,scale=0.26]{X2_OCARS_116.eps}
\centering \includegraphics[angle=-90,scale=0.26]{X1_OCARS_116.eps}
\centering \includegraphics[angle=-90,scale=0.26]{X0_OCARS_116.eps}
\caption{As Fig.~\ref{radio_plots} but for the OCARS quasars.}
\label{OCARS_plots}
\end{figure*}

In this appendix we test the potential of using the  OCARS quasars 
to find a radio photometric redshift model. In Fig.~\ref{OCARS_plots}, we show the redshift distributions of 
the features (Sect.~\ref{rfit}), from which the steep cut-off at lower flux densities, evident for the Milliquas QSOS, 
appears to be missing (cf. Fig.~\ref{radio_plots}). 
In addition to having higher fluxes, the correlations can also be quite different (Table~\ref{OCARS_stats}).
\begin{table}  
\begin{minipage}{55mm}
\centering
 \caption{The feature redshift correlations for the OCARS quasars  (Fig.~\ref{OCARS_plots}).}
\begin{tabular}{@{}l  r c r @{}}
\hline
\smallskip
Feature &  $n$ & $P(\tau)$  &  $Z(\tau)$\\ 
\hline
$\log_{10} S_{\text{70 MHz}}$  & 1137 & $2.69\times10^{-11} $& $6.67\sigma$ \\ 
$\log_{10} S_{\text{150 MHz}}$  & 800 & $2.69\times10^{-15} $& $7.91\sigma$ \\ 
$\log_{10} S_{\text{400 MHz}}$  & 1815 & $7.21\times10^{-16} $& $8.07\sigma$ \\
$\log_{10} S_{\text{700 MHz}}$  & 210 &  0.030& $2.17\sigma$\\ 
$\log_{10} S_{\text{1.0 GHz}}$  & 102 & 0.084& $1.73\sigma$ \\ 
$\log_{10} S_{\text{1.4 GHz}}$  &2908 & $4.48\times10^{-3}$& $2.82\sigma$ \\ 
$\log_{10} S_{\text{2.7 GHz}}$  & 1100 & $7.65\times10^{-9}$& $5.78\sigma$ \\ 
$\log_{10} S_{\text{5.0 GHz}}$  & 2966 & 0.047 & $1.99\sigma$ \\ 
$\log_{10} S_{\text{8.7 GHz}}$  & 2540 & 0.039 & $2.07\sigma$ \\ 
$\log_{10} S_{\text{15 GHz}}$ & 1016 & $5.80\times10^{-4}$& $3.43\sigma$  \\ 
$\log_{10} S_{\text{20 GHz}}$ & 1470 & $3.35\times10^{-11}$& $6.63\sigma$ \\
$\alpha_{1.4~\text{GHz}}$ & 1573 & 0.032 & $2.13\sigma$ \\ 
$|\alpha_{1.4~\text{GHz}}|$ & 1573 & 0.038 & $2.07\sigma$\\
$\log_{10} \nu_{\text{peak}}$ & 1169 & 0.915 & $0.11\sigma$\\
$\log_{10} S_{\text{peak}}$ & 1169 & 0.113& $1.59\sigma$ \\
$\alpha_{\text{thin}}$ & 1169 & 0.423 & $0.80\sigma$\\
$\alpha_{\text{thick}}$ & 1169 & 0.156 & $1.42\sigma$ \\ 
$\log_{10}\nu_{\text{TO}}$     & 1169 & $5.57\times10^{-5}$& $4.01\sigma$\\
$\log_{10}S_{\text{TO}}$     &1169 & $2.43\times10^{-4}$& $3.65\sigma$ \\
$\log_{10} (-X_2)$  & 1169&  $6.54\times10^{-3}$& $2.72\sigma$ \\
$\log_{10} X_1$  &1169 & 0.012 & $2.50\sigma$ \\
$\log_{10} (-X_0)$ & 1169 & 0.022 & $2.30\sigma$ \\
\hline
\end{tabular}
\label{OCARS_stats}  
\end{minipage}
\end{table} 
Specifically the strong redshift correlation with
$\log_{10} S_{\text{150 MHz}}$, $\log_{10} S_{\text{15 GHz}}$, $\log_{10} S_{\text{20 GHz}}$, $\log_{10}\nu_{\text{TO}}$ and 
$\log_{10}S_{\text{TO}}$, missing for the Milliquas sources (Table~\ref{z_stats}), but the   lack of strong correlation with 
$\log_{10} S_{\text{700 MHz}}$,  $\log_{10} S_{\text{1.0 GHz}}$ and $\log_{10} S_{\text{1.4 GHz}}$.

In Table~\ref{OCARS_overlap}, we also see that the distribution of the feature counts  is quite different from the Milliquas
sources (Table~\ref{intersect_all}), 
\begin{table}  
\begin{minipage}{67mm}
\centering
 \caption{Features common to  the sources in (column) descending order of sample size (Table~\ref{OCARS_stats}).}
\begin{tabular}{@{}l  r r | l  rr @{}}
\hline
\smallskip  
Feature &  $n$ & $n_{\text{int}}$ & Feature &  $n$ & $n_{\text{int}}$ \\ 
\hline
$\log_{10} S_{\text{5.0 GHz}}$  & 2966 &  2966  & \citeauthor{ssd+98} fit & 1169 & 311\\
$\log_{10} S_{\text{1.4 GHz}}$ & 2908 &    2772 &     $\log_{10} S_{\text{70 GHz}}$ & 1137 & 89\\      
$\log_{10} S_{\text{8.7 GHz}}$  & 2540 &  2281& $\log_{10} S_{\text{2.7 GHz}}$  & 1100 & 69\\
$\log_{10} S_{\text{400 MHz}}$  & 1815 &  1464 &  $\log_{10} S_{\text{15 GHz}}$ & 1016 & 55\\ 
$\alpha_{1.4~\text{GHz}}$ &  1573 &    706  &  $\log_{10} S_{\text{150 MHz}}$  &  800 & 30\\
$\log_{10} S_{\text{20 GHz}}$ & 1470 & 458&      $\log_{10} S_{\text{700 MHz}}$  & 210 & 20\\
Polynomial fit &1169 &  311 & $\log_{10} S_{\text{1.0 GHz}}$ &102 & 12\\
\hline
\end{tabular}
\label{OCARS_overlap}  
\end{minipage}
\end{table} 
although the inclusion of all features still results in a total sample size of just 12. Again,
prioritising by the most overlapping features, while still retaining a sufficient number of sources,
we test various machine learning models, equivalent to  those tested for the Milliquas
QSOs  (Sect.~\ref{rpr}). 
\begin{table*}  
\begin{minipage}{150mm}
\centering
 \caption{Results from the validation of the models (Sect.~\ref{rpr}), based upon the OCARS features (Table~\ref{OCARS_overlap}).}
\begin{tabular}{@{} l  c r c c r  c c  c @{}}
\hline
\smallskip
   & & \multicolumn{3}{c}{Un-normalised} & \multicolumn{4}{c}{Normalised}\\
  &  $r$   & $\mu_{\Delta z}$ & $\sigma_{\Delta z}$ & $\sigma_{\text{MAD}}$ &  $\mu_{\Delta z (\text{norm})}$ &  $\sigma_{\Delta z (\text{norm})}$ & $\sigma_{\text{NMAD}}$ & $\eta$ \\
\hline
 \multicolumn{9}{c}{$n_{\text{int}}\geq311$ of  the {\sc poly} sample  (63 validation sources)}\\
\hline
kNN & $0.19\pm0.12$ &  $-0.04\pm0.11$ &  $0.77\pm0.07$ &  $0.75\pm0.11$ &  $-0.12\pm0.05$ &  $0.38\pm0.04$ &  $0.31\pm0.05$ &  $60.5\pm6.0$\% \\
SVR & $0.20\pm0.10$ &  $0.03\pm0.10$ &  $0.77\pm0.07$ &  $0.71\pm0.12$ &  $-0.09\pm0.05$ &  $0.37\pm0.03$ &  $0.29\pm0.05$ &  $60.1\pm5.9$\%\\ 
DTR & $0.09\pm0.12$ &  $-0.03\pm0.12$ &  $0.86\pm0.08$ &  $0.79\pm0.13$ &  $-0.12\pm0.06$ &  $0.42\pm0.05$ &  $0.32\pm0.05$ &  $63.3\pm6.4$\% \\
ANN & $0.03\pm0.09$ &  $-0.03\pm0.32$ &  $0.85\pm0.05$ &  $0.84\pm0.13$ &  $-0.10\pm0.14$ &  $0.38\pm0.05$ &  $0.34\pm0.06$ &  $64.9\pm7.5$\%\\
\hline
\multicolumn{9}{c}{$n_{\text{int}}\geq311$ of the {\sc snellen} sample (63  validation sources)}\\
\hline
kNN & $0.10\pm0.10$ &  $0.00\pm0.12$ &  $0.79\pm0.06$ &  $0.77\pm0.11$ &  $-0.11\pm0.06$ &  $0.38\pm0.04$ &  $0.31\pm0.04$ &  $63.1\pm5.8$\%\\
SVR & $0.09\pm0.11$ &  $0.04\pm0.12$ &  $0.80\pm0.06$ &  $0.80\pm0.11$ &  $-0.09\pm0.06$ &  $0.39\pm0.04$ &  $0.32\pm0.04$ &  $65.0\pm5.2$\% \\
DTR & $0.00\pm0.11$ &  $-0.01\pm0.13$ &  $0.89\pm0.08$ &  $0.86\pm0.12$ &  $-0.12\pm0.06$ &  $0.44\pm0.06$ &  $0.35\pm0.05$ &  $66.7\pm5.5$\%\\
ANN & $0.01\pm0.07$ &  $0.03\pm0.27$ &  $0.86\pm0.04$ &  $0.91\pm0.08$ &  $-0.08\pm0.12$ &  $0.39\pm0.04$ &  $0.36\pm0.04$ &  $68.1\pm4.5$\%\\
\hline
\multicolumn{9}{c}{$n_{\text{int}}\geq706$ of the {\sc ohe} sample (142 validation sources)}\\
\hline   
kNN & $0.28\pm0.06$ &  $0.01\pm0.07$ &  $0.80\pm0.05$ &  $0.79\pm0.07$ &  $-0.10\pm0.03$ &  $0.38\pm0.03$ &  $0.32\pm0.03$ &  $61.7\pm3.6\%$  \\
SVR & $0.21\pm0.07$ &  $0.06\pm0.08$ &  $0.86\pm0.06$ &  $0.80\pm0.07$ &  $-0.08\pm0.03$ &  $0.41\pm0.04$ &  $0.33\pm0.03$ &  $63.4\pm3.5$\% \\ 
 DTR & $0.19\pm0.07$ &  $0.01\pm0.07$ &  $0.84\pm0.06$ &  $0.82\pm0.07$ &  $-0.10\pm0.03$ &  $0.40\pm0.03$ &  $0.33\pm0.03$ &  $64.1\pm3.7$\%\\
 ANN & $0.23\pm0.07$ &  $-0.16\pm0.29$ &  $0.80\pm0.04$ &  $0.91\pm0.15$ &  $-0.17\pm0.13$ &  $0.40\pm0.05$ &  $0.38\pm0.06$ &  $69.3\pm3.7$\%\\
\hline
\multicolumn{9}{c}{$n_{\text{int}}\geq458$ of the {\sc ohe} sample (92 validation sources)}\\
\hline   
kNN & $0.21\pm0.09$ &  $-0.02\pm0.08$ &  $0.74\pm0.06$ &  $0.72\pm0.09$ &  $-0.11\pm0.04$ &  $0.37\pm0.03$ &  $0.30\pm0.03$ &  $60.2\pm4.5$\% \\ 
SVR & $0.20\pm0.08$ &  $0.05\pm0.08$ &  $0.76\pm0.05$ &  $0.74\pm0.09$ &  $-0.08\pm0.04$ &  $0.37\pm0.03$ &  $0.30\pm0.04$ &  $59.8\pm4.8$\%\\ 
DTR & $0.12\pm0.11$ &  $-0.02\pm0.09$ &  $0.79\pm0.07$ &  $0.75\pm0.09$ &  $-0.11\pm0.04$ &  $0.40\pm0.05$ &  $0.30\pm0.04$ &  $61.7\pm4.7$\%\\
ANN & $0.14\pm0.07$ &  $-0.03\pm0.29$ &  $0.74\pm0.04$ &  $0.77\pm0.10$ &  $-0.10\pm0.13$ &  $0.38\pm0.05$ &  $0.32\pm0.05$ &  $62.1\pm6.0$\%\\ 
\hline
\multicolumn{9}{c}{$n_{\text{int}}\geq1464$ of the  {\sc flux} sample (293 validation sources)}\\
\hline 
 kNN &  $0.23\pm0.05$ &  $-0.02\pm0.05$ &  $0.75\pm0.04$ &  $0.75\pm0.04$ &  $-0.11\pm0.02$ &  $0.35\pm0.02$ &  $0.33\pm0.02$ &  $65.5\pm2.4$\% \\
 SVR & $0.27\pm0.04$ &  $0.09\pm0.05$ &  $0.74\pm0.04$ &  $0.70\pm0.04$ &  $-0.05\pm0.02$ &  $0.33\pm0.02$ &  $0.32\pm0.02$ &  $64.5\pm2.7$\% \\
DTR & $0.15\pm0.05$ &  $-0.01\pm0.05$ &  $0.77\pm0.04$ &  $0.77\pm0.05$ &  $-0.11\pm0.02$ &  $0.36\pm0.02$ &  $0.34\pm0.02$ &  $66.1\pm2.9$\%\\ 
ANN & $0.17\pm0.03$ &  $0.01\pm0.14$ &  $0.76\pm0.02$ &  $0.76\pm0.05$ &  $-0.09\pm0.07$ &  $0.37\pm0.03$ &  $0.34\pm0.02$ &  $66.3\pm1.4$\%\\
\hline
\multicolumn{9}{c}{$n_{\text{int}}\geq930$ of the  {\sc flux} sample (186 validation sources)}\\
\hline
kNN & $0.18\pm0.06$ &  $-0.03\pm0.05$ &  $0.70\pm0.04$ &  $0.72\pm0.06$ &  $-0.11\pm0.02$ &  $0.34\pm0.02$ &  $0.33\pm0.03$ &  $63.4\pm3.1$\%\\ 
SVR & $0.27\pm0.05$ &  $0.07\pm0.05$ &  $0.69\pm0.04$ &  $0.65\pm0.05$ &  $-0.06\pm0.02$ &  $0.32\pm0.02$ &  $0.31\pm0.02$ &  $62.0\pm3.0$\%\\ 
DTR & $0.13\pm0.06$ &  $-0.00\pm0.05$ &  $0.73\pm0.04$ &  $0.72\pm0.06$ &  $-0.10\pm0.02$ &  $0.35\pm0.02$ &  $0.33\pm0.03$ &  $63.9\pm3.1$\% \\
NZ & $0.27\pm0.04$ &  $0.11\pm0.11$ &  $0.78\pm0.02$ &  $0.70\pm0.06$ &  $-0.05\pm0.05$ &  $0.34\pm0.02$ &  $0.33\pm0.02$ &  $63.5\pm2.3$\%\\
\hline
\multicolumn{9}{c}{$n_{\text{int}}\geq531$ of the  {\sc flux} sample (107 validation sources)}\\
\hline
kNN & $0.08\pm0.08$ &  $-0.02\pm0.07$ &  $0.70\pm0.06$ &  $0.71\pm0.06$ &  $-0.11\pm0.04$ &  $0.34\pm0.02$ &  $0.33\pm0.03$ &  $63.0\pm4.0$\%\\ 
SVR & $0.21\pm0.08$ &  $0.09\pm0.07$ &  $0.69\pm0.06$ &  $0.65\pm0.07$ &  $-0.05\pm0.03$ &  $0.32\pm0.02$ &  $0.31\pm0.03$ &  $63.7\pm4.2$\%\\
DTR &  $0.09\pm0.10$ &  $0.00\pm0.08$ &  $0.74\pm0.06$ &  $0.71\pm0.08$ &  $-0.09\pm0.04$ &  $0.35\pm0.03$ &  $0.33\pm0.04$ &  $63.7\pm4.5$\\
ANN & $0.16\pm0.05$ &  $-0.02\pm0.15$ &  $0.74\pm0.02$ &  $0.70\pm0.06$ &  $-0.09\pm0.07$ &  $0.34\pm0.02$ &  $0.31\pm0.02$ &  $62.9\pm2.9$\%\\
\hline 
\end{tabular}
\label{OCARS_ML}  
\end{minipage}
\end{table*} 
However, from Table~\ref{OCARS_ML}, we see little promise of an accurate radio photometric redshift
with the results being poorer than for the Milliquas data  (Table~\ref{ML_TO}). It should be noted though that the training samples
are significantly smaller and so a poorer performance is consistent with our expectations (see Fig.~\ref{frac_test}).

\section*{Appendix B}
\label{appb} 
\subsection*{Comparison with published Milliquas photometric redshifts}

\begin{figure}
\centering \includegraphics[angle=-90,scale=0.5]{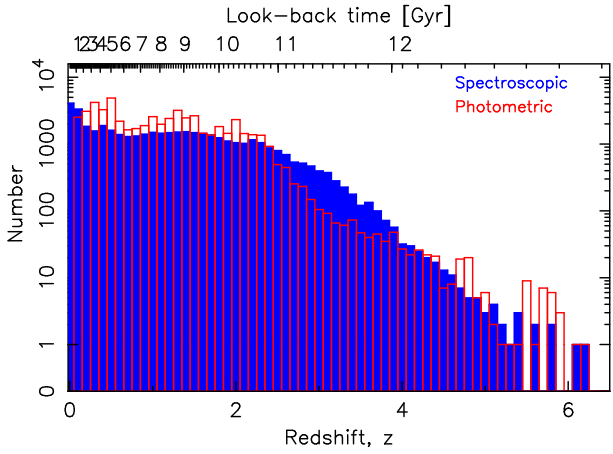}
\caption{The spectroscopic and published photometric redshift distributions of the Milliquas radio sources.
The latter are rounded to $0.1z$ \citep{fle21} and so there no photometric redshifts with $z_{\text {phot}}<0.1$. }
\label{z_phot}
\end{figure}

Of the 136\,076 Milliquas sources with a radio association, 57\,395 are quoted with
a photometric redshift \citep{fle21}\footnote{https://cdsarc.cds.unistra.fr/viz-bin/ReadMe/VII/290?format=html\&tex=true}, 
obtained using the four-colour method of \citet{fle15}. Of these, 56\,535 could be matched with a source in NED and 
in Fig.~\ref{z_phot} we show the distribution of the Milliquas radio sources with published photometric redshifts.

Again, scraping the NIR--optical--UV photometry from NED, WISE and GALEX (Sect.~\ref{opt_fit}), only 4844 of the 56\,535 sources have all nine magnitudes (Table~\ref{missing_phot}),
\setlength{\tabcolsep}{0.4em}
\begin{table}
\centering
\caption{The number of magnitude measurements in each of the bands for the 56\,535  Milliquas radio associated sources with photometric redshifts.}
\begin{tabular}{@{}c c c c c c c c c c @{}}
\hline
\smallskip 
$FUV$ &    $NUV $  & $u$ &  $g$ &    $r$  &     $i$ &   $z$ &    $W1$ &    $W2$ &    \\
12\,200 & 31\,326 & 22\,070& 24\,066 & 24\,543 & 29\,876 & 33\,394 & 54\,102 & 54\,454 \\
\hline
\end{tabular}
\label{missing_phot}  
\end{table} 
with the largest deficit being in the $FUV$. Given that the $NUV$ magnitude is detected in 55\% of the sources,
suggests that this may be a sensitivity issue. We could impute some of the missing magnitudes to increase
the sample size, but 4844 should be sufficient to compare with the Milliquas photometric redshifts without 
compromising our model.

Showing the redshifts we predict from our ANN (Sect.~\ref{stv})
\begin{figure} 
\centering \includegraphics[angle=-90,scale=0.5]{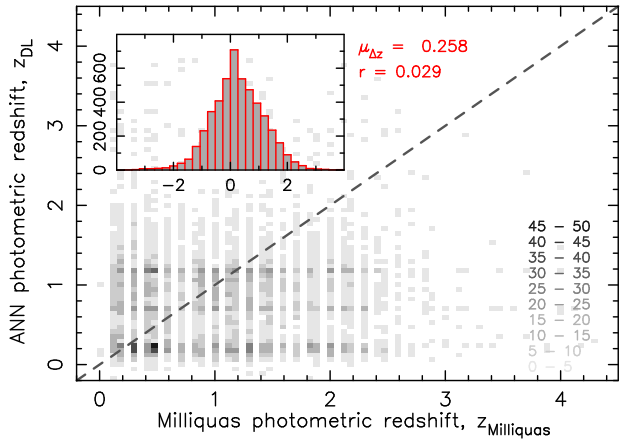}
\caption{The ANN photometric redshifts versus those of the Milliquas \citep{fle21}. 
$n=4844$, $\sigma_{\Delta z} = 0.92$,  $\mu_{\Delta z}=0.26$, $\sigma_{\text{MAD}}=0.90$, $\mu_{\Delta z (\text{norm})}=0.02$, $\sigma_{\Delta z(\text{norm})} = 0.47$, $\sigma_{\text{NMAD}}=0.45$ and $\eta=74$\%.} 
\label{Milliquas_photo}
\end{figure} 
against those of the Milliquas using the four-colour method (Fig.~\ref{Milliquas_photo}), suggests that these are 
generally unreliable. The four-colour method estimates photometric redshifts from the $u-g, g-r$, $r-i$ and $i-z$ colours
\citep{fle15}, but due to the shifting of rest-frame features into other observing bands, 
the NIR and UV colours (or magnitudes) are
crucial to obtaining accurate photometric estimates over a wide range of redshifts \citep{cur20}.

\end{document}